\documentclass[aps, prc, reprint]{revtex4-2}

\usepackage[breaklinks, colorlinks=true]{hyperref}
\usepackage{bm}
\usepackage{amsmath}
\usepackage{amssymb}
\usepackage{graphicx}
\usepackage{subcaption}

\newcommand{\unit}[1]{\,\mathrm{#1}}
\newcommand{\eqnl}{\nonumber\\&\qquad}

\begin{document}

%%%%%%%%%%%%%%%%%%%%%%%%%%%%%%%%%%%%%%%%%%%%%%%%%%%%%%%%%
\title{Constraining Energy Density Functionals via Bayesian Analysis of Nuclear Densities}
\date{\today}
\author{Chengpeng Yu}
\email{yucp@rcnp.osaka-u.ac.jp}
\affiliation{Research Center for Nuclear Physics, The University of Osaka, Ibaraki, Osaka 567-0047, Japan}
\author{Kenichi Yoshida}
\email{kyoshida@rcnp.osaka-u.ac.jp}
\affiliation{Research Center for Nuclear Physics, The University of Osaka, Ibaraki, Osaka 567-0047, Japan}
\affiliation{RIKEN Nishina Center for Accelerator-Based Science, Wako, Saitama 351-0198, Japan}
\affiliation{Center for Computational Sciences, University of Tsukuba, Tsukuba, Ibaraki 305-8577, Japan}
\begin{abstract}
    In nuclear many-body physics, energy density functional (EDF) theory is one of the most powerful approaches for describing finite nuclei and nuclear matter.
    However, its predictive capability depends on calibrating model parameters to experimental and observational data.
    In this work, we investigate an alternative approach: Constraining the parameters with the continuous density profiles of finite nuclei obtained from ab initio calculations.
    We apply Bayesian analysis to infer the parameters of Skyrme EDF from the density profiles and binding energies of $^{16}$O, $^{40}$Ca, and $^{48}$Ca.
    We show that the data effectively constrain the parameters associated with the properties of uniform nuclear matter, whereas those governing non-uniform nuclear matter remain partially constrained and require additional input.
    Furthermore, using the inferred parameter distributions, we successfully predict the density profiles and binding energy of $^{208}$Pb, which is excluded from the training data.
    This demonstrates the predictive capability of the framework.
    In conclusion, these results establish Bayesian analysis of density profiles as a promising route for incorporating accurate ab initio results of light nuclei into EDF development and strengthening the connection between both approaches.
\end{abstract}
\maketitle
%%%%%%%%%%%%%%%%%%%%%%%%%%%%%%%%%%%%%%%%%%%%%%%%%%%%%%%%%

%%%%%%%%%%%%%%%%%%%%%%%%%%%%%%%%%%%%%%%%%%%%
\section{Introduction}
%%%%%%%%%%%%%%%%%%%%%%%%%%%%%%%%%%%%%%%%%%%%%%%%
In nuclear many-body physics, energy-density functional (EDF) theory is among the most prominent approaches to establish a unified description of finite nuclei \cite{Bender:2003jk, Nakatsukasa:2016nyc, Colo:2020vik, Nakatsukasa2026} and neutron stars \cite{Yang:2019fvs, Nakatsukasa:2025axc, Yu:2025hmc, Chamel:2025fhb}.
EDF, the central component of this approach, is a functional that connects the system's energy to particle densities.
Currently, most EDFs are constructed based on symmetry and phenomenological considerations. They involve undetermined parameters.
Accurately calibrating the parameters has become a major challenge to improve the predictive power of EDF theory.

To calibrate the parameters, the community has been fitting EDF predictions to different experimental and observational constraints for specific scientific purposes throughout the history of nuclear physics. 
In the case of the well-established Skyrme EDF \cite{Vautherin:1971aw, Vautherin:1973zz, Yoshida:2023zaa, Minato:2025ozj}, the SkM* parameterization is calibrated on finite-nuclei and fission observables \cite{Bartel:1982ed, Dzhioev:2025wey, An:2025twk}, the SLy4 parameterization is calibrated on nuclear matter observables \cite{Chabanat:1997qh, Chabanat:1997un, Qu:2025vib}, and the more recent UNEDF parameterization is established by a global fit on finite-nucleus properties \cite{Kortelainen:2010hv, Kortelainen:2011ft, Kortelainen:2013faa}.
Interestingly, these parameterizations are mostly based on a finite set of scalar observables (e.g., binding energies, charge radii, separation energies, excitation energies, deformation parameters, fission barriers, etc.; see Ref.~\cite{Sun:2023xkg} for details).
In contrast, the spatially resolved neutron and proton density profiles of finite nuclei, $\rho_n(r)$ and $\rho_p(r)$, are considerably more difficult to determine experimentally and are seldom utilized to constrain EDFs \cite{Kurasawa:2020fli, Horiuchi:2021dku, Miyagi:2025lmv}.

In recent years, the field of nuclear theory has witnessed a rapid development of ab initio techniques (e.g., see Refs.~\cite{Barrett:2013nh, Launey:2021sua, Sun:2026eep} for the no-core shell model, Refs.~\cite{Hergert:2015awm, Heinz:2024juw, Miyagi:2025rvx} for the in-medium similarity renormalization group, and Refs.~\cite{Hiyama:2003cu, Hiyama:2012sma, Masui:2014nma, Aoyama:2026fbd} for the few-body Gaussian expansion method).
These techniques provide a promising avenue to acquire knowledge of the density profiles with moderate error bars.
For example, for well-converged light nuclei and in the nuclear interior,  the residual many-body uncertainty of the no-core shell model can be at the few-percent level \cite{Burrows:2017wqn, Burrows:2018ggt, Foy:2025yot, Sun:2025yfo}.
Accurate density profiles like this should provide powerful constraints on EDF parameters in complement to the conventional observables.
However, as far as we know, the impact has not been systematically investigated.

To quantify the impact of density profiles, one needs to analyze how they affect the uncertainties of EDF parameters and observables of interest.
This is challenging with conventional EDF calibration techniques, which are primarily point-like estimations of the parameters based on $\chi^2$ tests, implemented by gradient-based \cite{Hascoet:2025zfm, Yoshimura:2026cdk}, derivative-free \cite{Kortelainen:2013faa, Scamps:2020fyu}, or stochastic \cite{Agrawal:2005ix, Zhang:2015vaa, Amiri:2026zbz} optimization algorithms.
On the other hand, Bayesian analysis has been considered a powerful alternative to keep track of the uncertainties.
As a highly flexible machine learning framework, it combines physically informed assumptions with data input to propose a probability distribution of the quantity of interest, based on which uncertainty estimation becomes straightforward.
The Bayesian approach has been applied in different fields of nuclear physics for years (see Ref.~\cite{Phillips:2020dmw} for a review).
In the recent decade, it has been widely utilized to infer the neutron-star equation of state from the mass and radius observed in astrophysical projects like NICER (see Refs.~\cite{Miller:2019cac, Burgio:2021vgk, Choudhury:2024xbk} for the project and Refs.~\cite{Ozel:2016oaf, Semposki:2025etb, Pal:2026cji} for the Bayesian equation of state).
Because the equation of state is closely related to EDF, the community is increasingly interested in applying Bayesian analysis to constrain parameters of both Skyrme EDF \cite{Klausner:2024jgu, Klausner:2025ucq, Klausner:2026foh} and covariant EDF \cite{Neufcourt:2018syo, Li:2025oxi, Wei:2025aku, Xie:2026rep}.

However, although Bayesian analysis of EDF parameters has attracted growing interest, applying it to finite-nucleus observables faces a major challenge.
The reason is:
For uniform nuclear matter, many observables can be expressed explicitly as functions of the EDF parameters;
In contrast, observables associated with nonuniform nuclear systems generally require solving self-consistent equations to determine the underlying spatial structure.
Since Bayesian parameter estimation requires repeated evaluations over a broad parameter space, incorporating finite-nucleus calculations into the inference becomes expensive.
Currently, only a few works exist in this direction.
One representative work is the series of publications by P. Klausner \textit{et al.}; see Refs.~\cite{Klausner:2024jgu, Klausner:2025ucq, Klausner:2026foh}, in which the computation cost is reduced by applying a Gaussian-process emulator of the conventional EDF solver.
The works, however, focused on experimental observables of relatively heavy nuclei (e.g., $^{208}$Pb, $^{48}$Ca, $^{68}$Ni, $^{132}$Sn, $^{90}$Zr in Ref.~\cite{Klausner:2026foh}).
Most of the nuclei are beyond the reach of ab initio calculations in the near future.
Furthermore, to constrain the isovector part of the EDF, the works used experimental observables including polarizations and parity-violating asymmetries.
Although related, the information introduced by these observables is different from that by continuous neutron density profiles.
To our knowledge, no work has included the ab initio density profiles as the optimization target of Bayesian analysis of EDF so far. 

To fill this gap, this work seeks the possibility to constrain the parameters of the Skyrme EDF by a Bayesian analysis combining neutron and proton density profiles and binding energies of finite nuclei.
Specifically, this work takes $\rho_n(r)$, $\rho_p(r)$, and $B$ of $^{16}$O, $^{40}$Ca, and $^{48}$Ca (which are reachable by no-core shell model methods in the near future; see Refs.\cite{Burrows:2017wqn, Burrows:2018ggt, Foy:2025yot}) as input and computes the probability distribution of the Skyrme parameters.
Instead of using an emulator, this work uses a conventional EDF solver, HFBRAD, to compute the likelihood.
The obtained parameter distributions are applied to predict the density profiles and binding energy of $^{208}$Pb---a nucleus outside the training dataset and the reach of conventional ab initio methods in the foreseeable future---validating the predictive capability of the Bayesian framework.

The paper is organized as follows:
Sec.~\ref{sec:review} reviews the Skyrme EDF and the method to solve the density profiles of finite nuclei;
Sec.~\ref{sec:framework} describes the Bayesian framework that this work implements to infer the parameters:
Sec.~\ref{sec:results} presents the parameters distributions and predictions;
Finally, Sec.~\ref{sec:concl} summarizes the outcomes.

%%%%%%%%%%%%%%%%%%%%%%%%%%%%%%%%%%%%%%%%%%%%%%
\section{Skyrme EDF}
\label{sec:review}
%%%%%%%%%%%%%%%%%%%%%%%%%%%%%%%%%%%%%%%%%%%%%
In this section, we introduce the Skyrme EDF, which we plan to constrain.

The key assumption of the Skyrme EDF is that the energy of the system only depends on the local part of the density matrix $\rho(\bm{r}\sigma, \bm{r}'\sigma')$.
For the even-even nuclei we work on, the relevant local densities are the normal density, the kinetic density, and the spin-orbit current, defined as below,
\begin{equation}
    \rho(\bm{r}) = \sum_{\sigma} \rho(\bm{r}\sigma, \bm{r}\sigma),
\end{equation}
\begin{equation}
    \tau(\bm{r}) = \sum_{\sigma} \left. \nabla\cdot\nabla' \rho(\bm{r}\sigma, \bm{r}'\sigma) \right|_{\bm{r}=\bm{r}'},
\end{equation}
\begin{equation}
    \bm{J}(\bm{r}) = \sum_{\sigma\sigma'} \frac{i}{2} \left[ (\nabla'-\nabla) \times \left(\rho(\bm{r}\sigma, \bm{r}'\sigma') \bm{\sigma}_{\sigma'\sigma} \right) \right]_{\bm{r}=\bm{r}'}.
    \label{eq:J-so}
\end{equation}
Here, $\sigma$, $\sigma'$ are spin indices and $\bm{\sigma}_{\sigma'\sigma}$ is the matrix elements of the Pauli matrices.
The isospin indices are included implicitly.

From these local densities, one can formulate the Skyrme EDF of even-even nuclei as 
\begin{align}
    &\mathcal{E}[\rho] = \sum_{t=0,1} \biggl[
    \left( C^{\rho}_{t0} + C^{\rho}_{tD} \rho_0^{\gamma} \right) \rho_t^2 
    + C_t^{\Delta\rho} \rho_t \Delta \rho_t\eqnl
    + C_t^{\tau} \rho_t \tau_t
    + C_t^{\nabla J} \rho_t \nabla\cdot\bm{J}_t + \frac{1}{2}C_t^J \bm{J}_t^2\biggr].
    \label{eq:Skyrme}
\end{align}
Here, $t=0,1$ labels the isoscalar and isovector components of densities, i.e., $\rho_{0} = \rho_n + \rho_p$. $\rho_{1} = \rho_n - \rho_p$, with $\rho_n$ and $\rho_p$ being the neutron and proton densities, respectively.

In Eq.~\eqref{eq:Skyrme}, the term proportional to $\bm{J}_t^2$ has a small effect on finite-nucleus observables and is discarded in the SkM* and SLy4 parameterizations used in this work (see Refs.~\cite{Bennaceur:2005mx, Fracasso:2007fi} for discussions). Hence, we choose $C_t^J=0$.
Furthermore, we only consider the conventional spin-orbit interactions.
As discussed in Ref.~\cite{Klausner:2024jgu}, we fix $C_0^{\nabla J}=3C_1^{\nabla J}$.
Hence, the free Skyrme parameters in our setup are
\begin{align}
    & \theta = \biggl(
    C^{\rho}_{00},\, C^{\rho}_{0D},\, C_0^{\Delta\rho},\, C_0^{\tau},\, C_0^{\nabla J},\,\eqnl 
    C^{\rho}_{10},\, C^{\rho}_{1D},\, C_1^{\Delta\rho},\, C_1^{\tau},\, 
    \gamma\biggr).
    \label{eq:isospin-para}
\end{align}

In addition to the Skyrme EDF that depends on conventional density matrices, the energy of a nuclear system also depends on the density of Cooper pairs, $\kappa(\bm{r}\sigma,\bm{r}'\sigma') =  \langle \psi(\bm{r}'\sigma') \psi(\bm{r\sigma}) \rangle$.

Once the total energy density $E[\rho,\kappa]$ is determined, one can minimize it with respect to $\rho$ and $\kappa$ to compute ground-state observables.
Numerically, this is implemented by solving the Hartree-Fock-Bogoliubov equation,
\begin{equation}
    [H_{\text{HFB}},R]=0,\label{eq:HFB}
\end{equation}
iteratively with
\begin{equation}
    H_{\text{HFB}}=\frac{\delta}{\delta R^{T}}\left[E-\mu_{n}\left(\text{tr}\rho_{n}-N\right)-\mu_{p}\left(\text{tr}\rho_{p}-Z\right)\right],\label{eq:H_HFB}
\end{equation}
where the generalized density matrix $R$ is
\begin{equation}
    R=\left(\begin{array}{cc}
        \rho & \kappa\\
        -\kappa^{*} & 1-\rho^{*}
    \end{array}\right),\label{eq:R}
\end{equation}
$N$ and $Z$ represent the neutron and proton numbers, and $\mu_n$ and $\mu_p$ are the corresponding Lagrange multipliers.

In this work, the nuclei we are considering, $^{16}$O, $^{40}$Ca, $^{48}$Ca, and $^{208}$Pb, are double-magic nuclei.
The pairing contribution $\kappa$ vanishes.
However, it is straightforward to extend the Bayesian framework to include the pair contributions.

%%%%%%%%%%%%%%%%%%%%%%%%%%%%%%%%%%%%%%%%%%%%%%
\section{Bayesian framework}
\label{sec:framework}
%%%%%%%%%%%%%%%%%%%%%%%%%%%%%%%%%%%%%%%%%%%%%
In this section, we establish the Bayesian framework to determine the Skyrme parameters from the density profiles and binding energies of $^{16}$O, $^{40}$Ca, and $^{48}$Ca.

%%%%%%%%%%%%%%%%%%%%%%%%%%%%%%%%%%%%%%%
\subsection{Representation of data and parameters}
\label{ssec:space}

Since $^{16}$O, $^{40}$Ca, and $^{48}$Ca are spherically symmetric, the input data of density profiles only depend on radius.
We introduce a lattice $\{r_0, r_1,r_2,\cdots, r_{M-1}\}$ and represent the densities as
\begin{equation}
    \rho_{q}^{(N, Z)}=\left(\rho_{q}^{(N, Z)}(r_0),\, \rho_{q}^{(N, Z)}(r_1),\, \cdots,\, \rho_{q}^{(N, Z)}(r_{M-1})\right)
\end{equation}
with $q=n,\,p$ and $(N,Z)=(8,8),\,(20,20),\,(28,20)$.
The densities and binding energies should have error bars originating from, for example, the uncertainty of chiral effective theory and of the truncation introduced by ab initio methods.
To account for the uncertainties, we introduce two scalar variables $\sigma_{\rho}$ and $\sigma_B$ and formally represent the input data as
\begin{equation}
    X = \left(\rho_{q}^{(N,Z)},\, B^{(N,Z)},\, \sigma_\rho,\, \sigma_B\right).
\end{equation}
We will introduce the details of $\sigma_{\rho}$ and $\sigma_B$ in Sec.~\ref{ssec:likelihood}.

For each parameter $\theta$, we use Eqs.~\eqref{eq:HFB}--\eqref{eq:R} to predict the density profiles $\rho_{q}^{(N, Z)}(\theta)$ and binding energies $B^{(N,Z)}(\theta)$.
In spherically symmetric nuclei, a well-established open-source code, HFBRAD, has implemented this procedure; see Ref.~\cite{Bennaceur:2005mx} for details.
We use the code in this work.
Furthermore, one needs to consider that due to  1) the mean-field approximation, 2) the assumptions of the Skyrme EDF, and 3) the systematic errors of the iterative solver, even the EDF calculations with the best-fitting parameters may not be able to reproduce the input data within their quoted uncertainties.
To quantify the potential model discrepancy, we introduce the scalar variables $\sigma_{\rho}'$ and $\sigma_B'$.
While $\sigma_{\rho}$ and $\sigma_B$ are part of the input data, $\sigma_{\rho}'$ and $\sigma_B'$ are considered part of the parameter space and will be determined by the Bayesian analysis.
Hence, the parameters are formally represented as
\begin{equation}
    \vartheta = (\theta, \sigma_\rho', \sigma_B').
\end{equation}
We will introduce the details of $\sigma_{\rho}'$ and $\sigma_B'$ in Sec.~\ref{ssec:likelihood}.

The posterior distribution is
\begin{equation}
    P(\vartheta|X) = \frac{1}{Z} P(X|\vartheta) P(\vartheta),
    \label{eq:Baye}
\end{equation}
\begin{equation}
    Z = \int_{\vartheta} P(X|\vartheta) P(\vartheta).
    \label{eq:Z}
\end{equation}
Here, $P(X|\vartheta)$ and $P(\vartheta)$ are the likelihood and prior, respectively.
One can obtain the posterior distribution of $\theta$ by integrating out $\sigma_{\rho}'$ and $\sigma_B'$ from $P(\vartheta|X)$.

%%%%%%%%%%%%%%%%%%%%%%%%%%%%%%%%%%
\subsection{Likelihood}
\label{ssec:likelihood}
We define the likelihood as a Gaussian distribution:
\begin{align}
    & P(X|\vartheta)=\prod_{N,Z}\biggl( \frac{1}{\sqrt{2\pi(\sigma_{B}^{2}+\sigma_{B}^{\prime2})}}\frac{1}{2\pi(\sigma_{\rho}^{2}+\sigma_{\rho}^{\prime2})}\eqnl
    \qquad \times e^{-\frac{1}{2\sqrt{\sigma_{B}^{2}+\sigma_{B}^{\prime2}}}\left|B^{(N,Z)}-B^{(N,Z)}(\theta)\right|^{2}}\eqnl
    \qquad \times e^{-\frac{1}{2\sqrt{\sigma_{\rho}^{2}+\sigma_{\rho}^{\prime2}}}\sum_{q=n,p}\left\Vert \rho_{q}^{(N,Z)}-\rho_{q}^{(N,Z)}(\theta)\right\Vert ^{2}}\biggr).
    \label{eq:likelihood}
\end{align}
In the above equation, we treat $\sigma_{\rho}$, $\sigma_B$ and $\sigma_{\rho}'$, $\sigma_B'$ as independent noises since they are of different origins.
We define the norm between $\rho_{q}^{(N, Z)}$ and $\rho_{q}^{(N, Z)}(\theta)$ as
\begin{align}
    &\left\Vert \rho_{q}^{(N,Z)}-\rho_{q}^{(N,Z)}(\theta)\right\Vert \eqnl
    =\frac{\sum_{i}\left|\rho_{q}^{(N,Z)}(r_{i})-\rho_{q}^{(N,Z)}(r_{i};\theta)\right|\rho_{q}^{(N,Z)}(r_{i})r_{i}^{2}}{\sum_{i}\rho_{q}^{(N,Z)}(r_{i})r_{i}^{2}}.
\end{align}
In the definition, we introduce a weighted sum of $|\rho_{q}^{(N,Z)}(r_{i})-\rho_{q}^{(N,Z)}(r_{i};\theta)|$ with respect to the number of nucleons around $r_{i}$.
Compared with the naive unweighted sum, this definition forces the Bayesian framework to reproduce the behavior of densities in the surface region of nuclei, which relates to observables of interest such as the neutron skin thickness.

Interestingly, in the above definition of likelihood, the density part corresponds to a Gaussian process with kernel function
\begin{equation}
    \left(K^{(N,Z)}_t\right)^{-1}_{ij} = \frac{s_{i}s_{j}}{\sqrt{\sigma_{\rho}^{2}+\sigma_{\rho}'^{2}}}\frac{\rho_{q}^{(N,Z)}(r_{i})r_{i}^{2}\rho_{q}^{(N,Z)}(r_{j})r_{j}^{2}}{\left(\sum_{k}\rho_{q}^{(N,Z)}(r_{k})r_{k}^{2}\right)^{2}},
\end{equation}
with $s_i$ being the sign of $\rho_{q}^{(N,Z)}(r_{i})-\rho_{q}^{(N,Z)}(r_{i};\theta)$.
Therefore, the likelihood considers the correlation among the densities at different spatial locations. 

%%%%%%%%%%%%%%%%%%%%%%%%%%%%%%%%
\subsection{Prior}
\label{ssec:prior}
We assume independent priors of the Skyrme parameters and of the uncertainties, i.e., $P(\vartheta)=P(\theta) P(\sigma_\rho',\sigma_B')$.

We notice that the Skyrme parameters, $\theta = (C^{\rho}_{00},\, C^{\rho}_{0D},\, C_0^{\Delta\rho},\, C_0^{\tau},\, C_0^{\nabla J},\, C^{\rho}_{10},\, C^{\rho}_{1D},\, C_1^{\Delta\rho},\, C_1^{\tau},\, \gamma)$, are introduced as a compact formulation that facilitates theoretical derivations and isospin symmetry analysis, but each parameter alone does not have an evident physical interpretation.
Hence, it is challenging to propose a prior based on physical intuitions.
Instead of directly proposing $P(\theta)$, we transform the parameters into a different representation.
The new representation is based on the following properties of uniform and non-uniform nuclear matter (see \cite{Bender:2003jk} for detailed definitions):
\begin{itemize}
    \item Phenomenological equation of state of uniform nuclear matter, defined as 
    \begin{align}
        &\qquad e(\rho_{0},\beta)=\left[E_{0}+\frac{1}{2}K_{0}\left(\frac{\rho_{0}-\rho_{\text{sat}}}{3\rho_{\text{sat}}}\right)^{2}\right] \eqnl
        \qquad+ \biggl[J+L\left(\frac{\rho_{0}-\rho_{\text{sat}}}{3\rho_{\text{sat}}}\right)\eqnl
        \qquad\qquad+K_{\text{sym}}\left(\frac{\rho_{0}-\rho_{\text{sat}}}{3\rho_{\text{sat}}}\right)^{2}\biggr]\beta^{2},
        \label{eq:eos}
    \end{align}
    where $e(\rho_{0},\beta)$ is the energy per particle, $\beta=\rho_1/\rho_0$, and this equation is valid for small $\beta$ and $\rho_0-\rho_{\text{sat}}$.
    The parameter $\rho_{\text{sat}}$ is  the nuclear saturation density defined such that $\partial e(\rho_0,0)/\partial \rho_0 = 0$.
    Parameters $E_0$ and $K_0$ are the ground-state energy and incompressibility of symmetric nuclear matter, respectively.
    Parameters $J$, $L$, and $K_{sym}$ characterize the symmetry energy.
    \item Effective masses, defined such that
    \begin{equation}
        \frac{\hbar^2}{2m^*_{n,p}} = \frac{\delta}{\delta \tau_{n,p}} \biggl( 
            \frac{\hbar^2 \tau_{n,p}}{2m} + \mathcal{E}[\rho]
        \biggr),
        \label{eq:effmass-m0}
    \end{equation}
    where $m$ is the bare nucleon mass.
    When $\rho_n=\rho_p$, $m^*_n=m^*_p$. This mass is defined as the isoscalar effective mass $m_0^*$.
    On the other hand, the isovector effective mass is defined such that
    \begin{equation}
        \frac{\hbar^2}{2m_0^*} -\frac{\hbar^2}{2m_1^*} =  \frac{1}{2} \frac{\partial}{\partial\beta} \left(\frac{\hbar^2}{2m_n^*} - \frac{\hbar^2}{2m_p^*}\right).
        \label{eq:effmass-m1}
    \end{equation} 
    Throughout this manuscript, we use effective masses defined at $\rho_0=\rho_{\text{sat}}$ and $\beta=0$.
    \item The spin-orbit coupling term of the conventional Skyrme force, defined by
    \begin{equation}
        \hat{v}_{SO} = i W_0 (\bm{\sigma}_1+\bm{\sigma}_2) \cdot \left(\bm{k}^\dagger \times \delta(\bm{r}_{1}-\bm{r}_2) \bm{k}\right),
        \label{eq:interaction-so}
    \end{equation}
    with $\bm{k}=(-i/2)(\nabla_1-\nabla_2)$ and $W_0$ being the spin-orbit coupling constant.
\end{itemize}
One can show that there is a one-to-one mapping $f$ from $\theta$ to
\begin{equation}
    \Theta = \left(\rho_{\text{sat}},\, E_0,\, K_0,\, J,\, L,\, m_0^*,\, m_1^*,\, W_0,\, C_0^{\Delta\rho},\, C_1^{\Delta\rho}\right).
\end{equation}
For detailed derivation, see Appendix~\ref{app:trans}.
Here, both $\theta$ and $\Theta$ include $C_t^{\Delta\rho}$ since its physical interpretation---the density-gradient effect of finite nuclei---is clear.
Alternatively, one can parameterize the effect using the isoscalar and isovector surface-energy coefficients, $G_0$ and $G_1$; see Refs.~\cite{Chen:2009wv, Chen:2010qx}. 
Furthermore, $\Theta$ does not include $K_{\text{sym}}$ due to the limited number of free parameters in the present form of Skyrme EDF.
If necessary, one can include $K_{\text{sym}}$ and higher-order derivatives of the symmetry energy into the parameter space by generalizing the Bayesian framework to extended Skyrme EDF approaches, such as the KIDS functional, which has multiple density dependence~\cite{Gil:2020wqs}.
This will provide a better description of symmetric and neutron matter over different density regimes.

\begin{table}
  \caption{Uniform prior of Skyrme parameters}
  \label{tab:prior}
  \begin{ruledtabular}
    \begin{tabular}{ccccc}
      Name & Value \\
      \hline
      $\rho_{\text{sat}}\unit{(fm^{-3})}$  & $0.150$--$0.175$\\
      $E_0\unit{(MeV)}$  & $-16.50$--$-15.50$\\
      $K_0\unit{(MeV)}$  & $180.00$--$260.00$\\
      $J\unit{(MeV)}$   &   $25.00$--$40.00$\\
      $L\unit{(MeV)}$   &   $10.00$--$120.00$\\
      $C^{\Delta\rho}_0\unit{(MeV\cdot fm^5)}$    &   $-120.00$--$-20.00$\\
      $C^{\Delta\rho}_1\unit{(MeV\cdot fm^5)}$    &   $-50.00$--$50.00$\\
      $W_0\unit{(MeV\cdot fm^5)}$ &   $80.00$--$170.00$\\
      $m_0^*$   &   $(0.60$--$1.10)\,m$\\
      $m_1^*$   &   $(0.60$--$1.10)\,m$\\
    \end{tabular}
  \end{ruledtabular}
\end{table}

The existence of $f$ allows us to determine $P(\theta)$ from $P(\Theta)$.
For $\Theta$, we propose a uniform prior distribution for each parameter as shown in Tab.~\ref{tab:prior}.
The ranges are based on the analysis in Refs.~\cite{Kortelainen:2010hv, Zhao:2022xhq, Klausner:2024jgu} and the calculation of available Skyrme EDF parameterizations.

For the $\sigma_{\rho}'$, $\sigma_{B}' $ part of the prior, we propose a half-Gaussian distribution,
\begin{equation}
    P(\sigma_{\rho}',\sigma_{B}')=\frac{2}{\pi\bar{\sigma}_{\rho}\bar{\sigma}_{B}}e^{-\frac{1}{2}\left(\frac{\sigma_{\rho}^{\prime2}}{\bar{\sigma}_{\rho}^{2}}+\frac{\sigma_{B}^{\prime2}}{\bar{\sigma}_{B}^{2}}\right)},\qquad\sigma_{\rho}^{\prime},\sigma_{B}^{\prime}>0.
\end{equation}
We choose $\bar{\sigma}_{\rho} = 0.1\rho_{\text{sat}}$, $\bar{\sigma}_B=1\unit{MeV}$.

%%%%%%%%%%%%%%%%%%%%%%%%%%%%%%%%%%%%%%%
\subsection{Method to compute the posterior}
\label{ssec:method}

%%%%%%%%%%%%%%%%%%%%%%%%%%%%%%%%%%%%%%%%%%%%%%%%%%%
\begin{figure*}
    \centering
    \includegraphics[width=\linewidth]{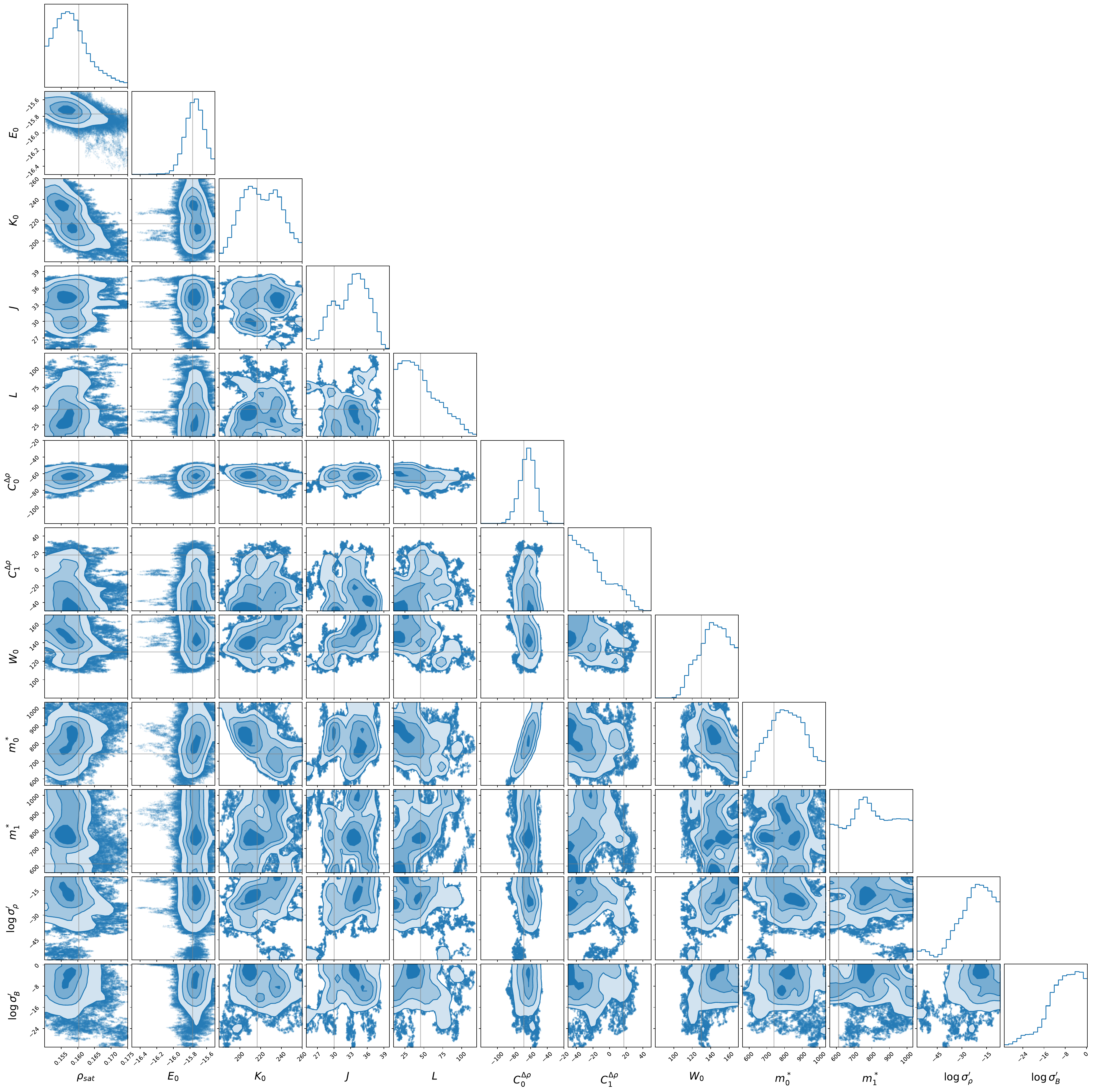}
    \caption{Corner plot of the Skyrme parameters. The EDF parameterization is SkM*. The input noises are $\sigma_{\rho}=0.01\rho_{\text{sat}}$, $\sigma_B = 0.30\unit{MeV}$. The gray lines represent the accurate values of SkM*. This figure is based on the Python package corner in Ref.~\cite{corner}.
}
    \label{fig:corner}
\end{figure*}
%%%%%%%%%%%%%%%%%%%%%%%%%%%%%%%%%%%%%%%%%%%%%%%%%%%

%%%%%%%%%%%%%%%%%%%%%%%%%%%%%%%%%%%%%%%%%%%%%%%%%%%
\begin{figure*}
    \centering
    \begin{subfigure}{0.32\linewidth}
        \includegraphics[width=\linewidth]{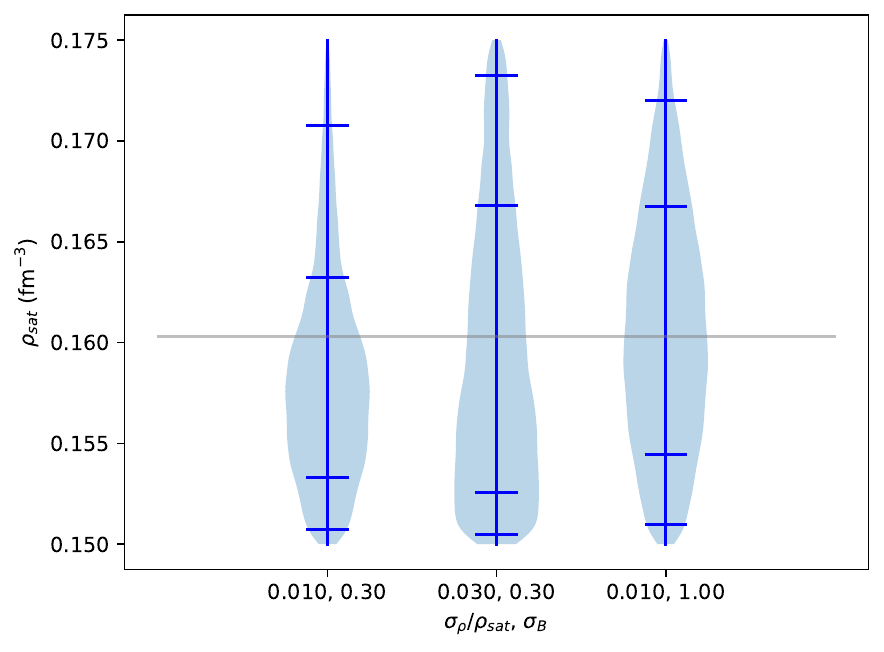}
    \end{subfigure}
    \begin{subfigure}{0.32\linewidth}
        \includegraphics[width=\linewidth]{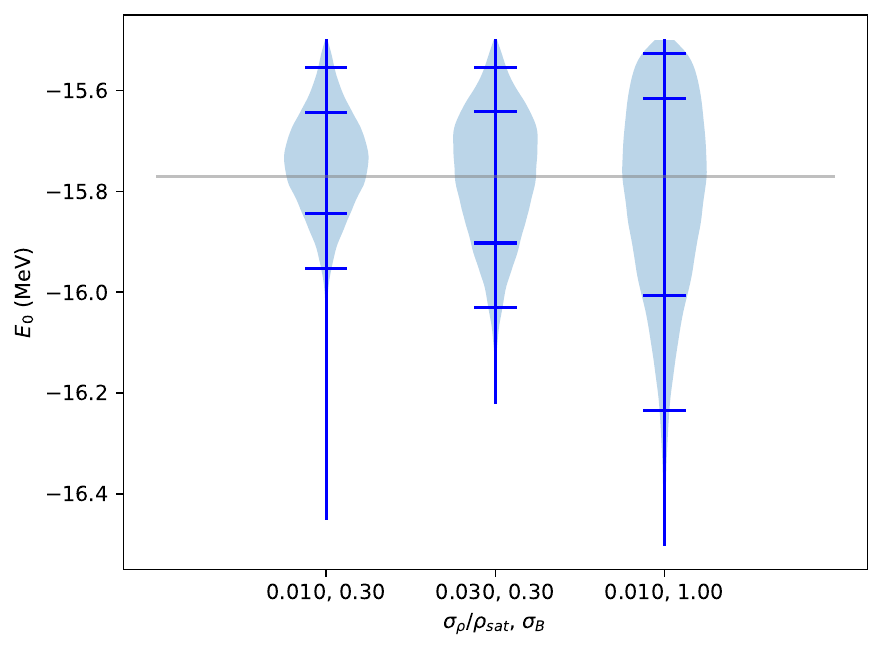}
    \end{subfigure}\\
    \begin{subfigure}{0.32\linewidth}
        \includegraphics[width=\linewidth]{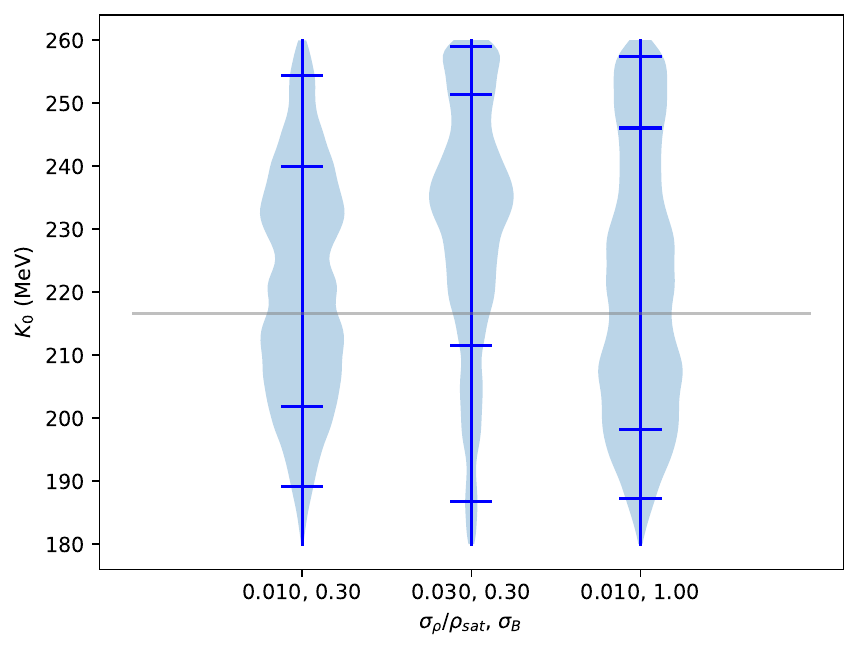}
    \end{subfigure}
    \begin{subfigure}{0.32\linewidth}
        \includegraphics[width=\linewidth]{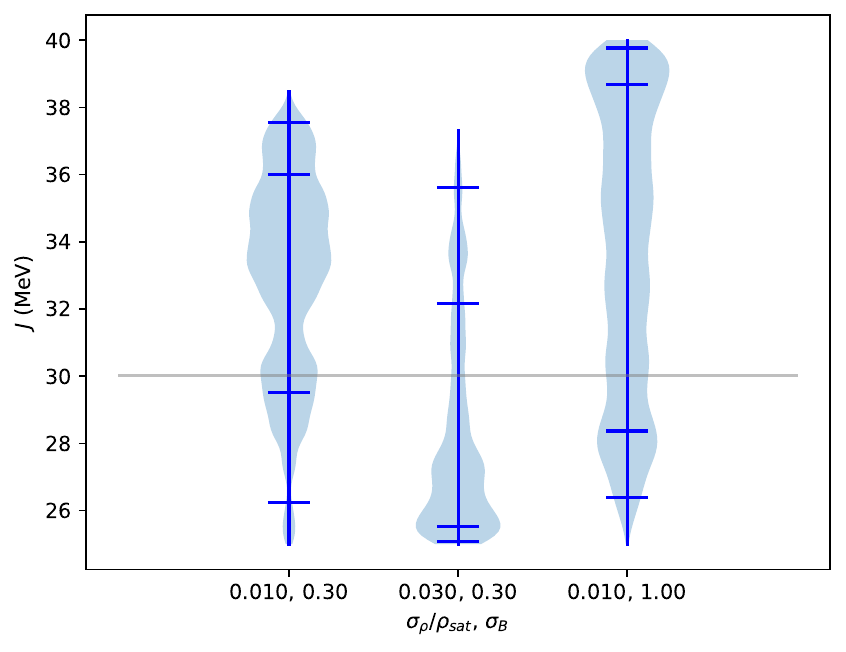}
    \end{subfigure}\\
    \begin{subfigure}{0.32\linewidth}
        \includegraphics[width=\linewidth]{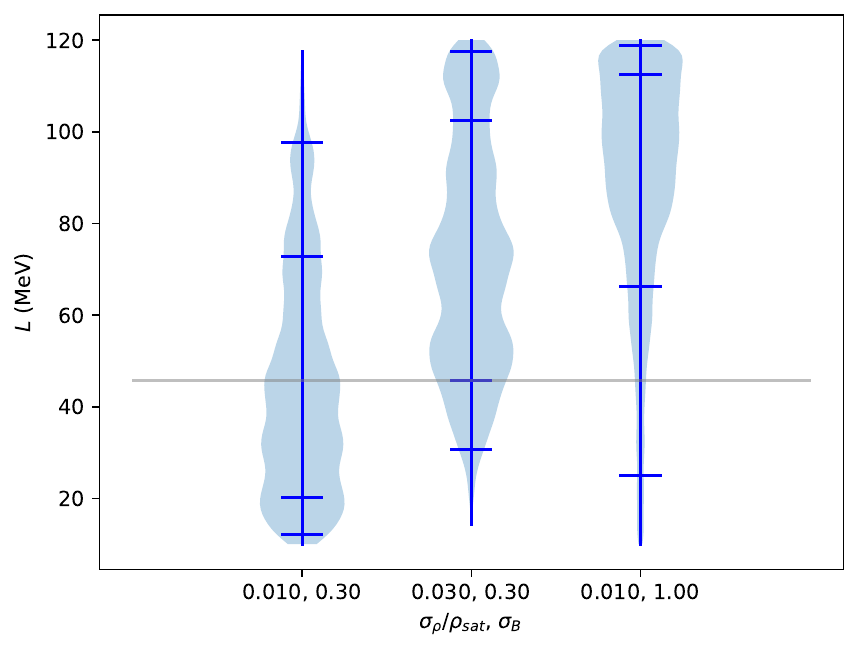}
    \end{subfigure}
    \begin{subfigure}{0.32\linewidth}
        \includegraphics[width=\linewidth]{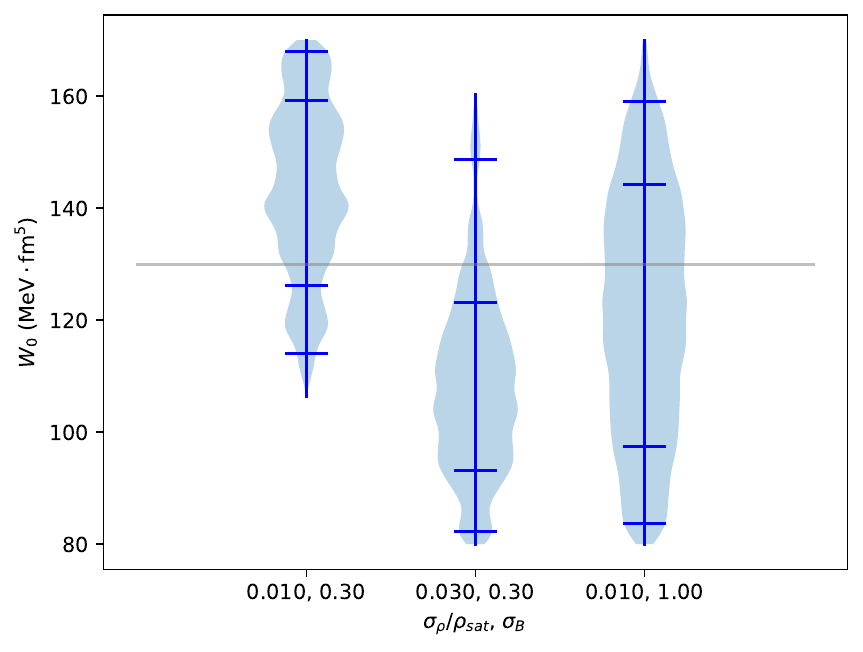}
    \end{subfigure}\\
    \begin{subfigure}{0.32\linewidth}
        \includegraphics[width=\linewidth]{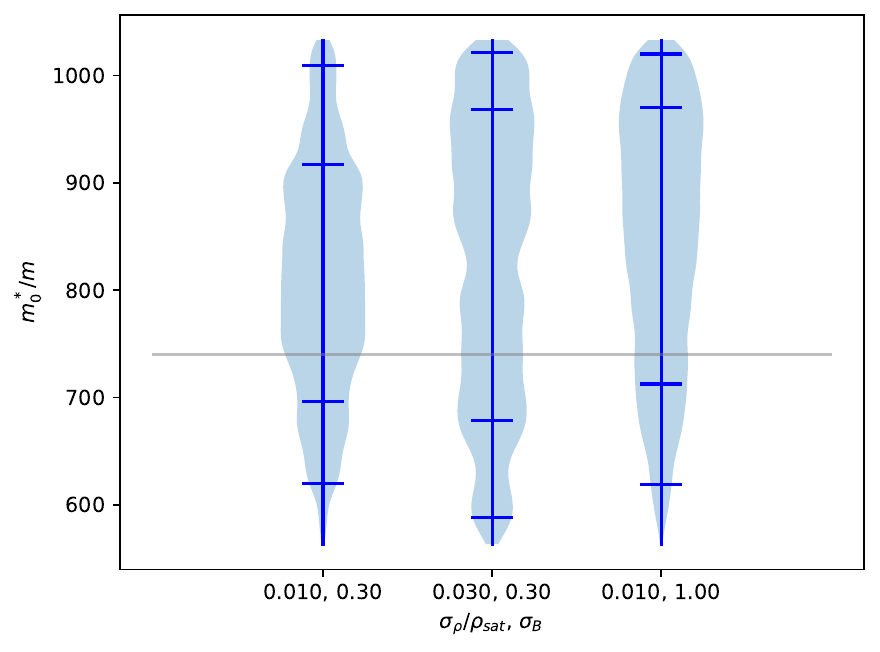}
    \end{subfigure}
    \begin{subfigure}{0.32\linewidth}
        \includegraphics[width=\linewidth]{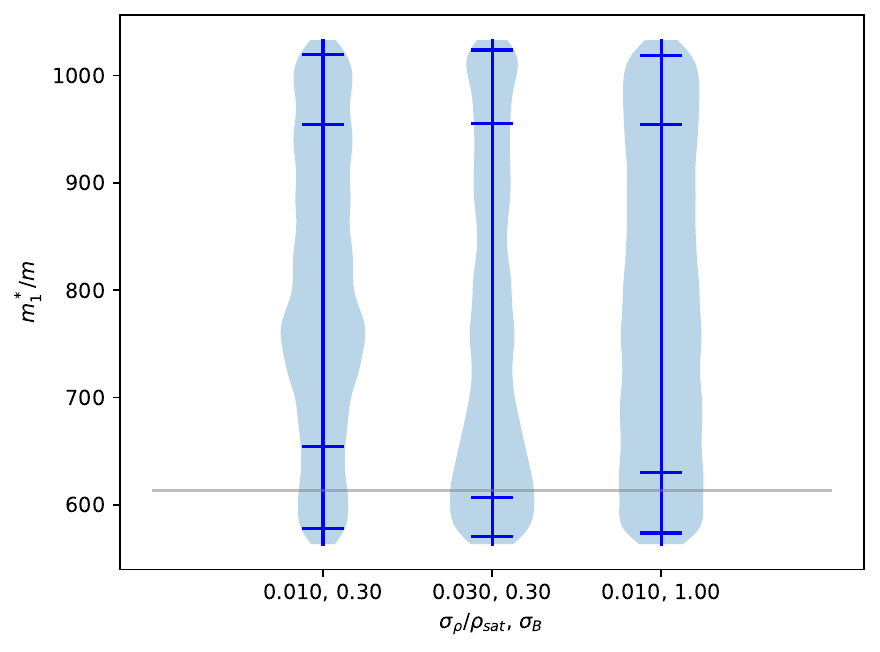}
    \end{subfigure}\\
    \begin{subfigure}{0.32\linewidth}
        \includegraphics[width=\linewidth]{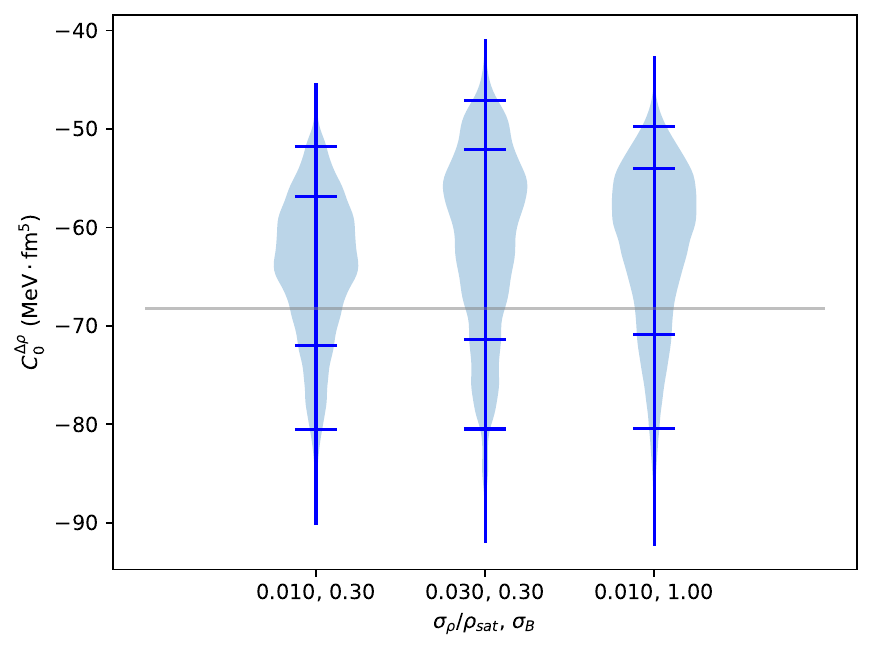}
    \end{subfigure}
    \begin{subfigure}{0.32\linewidth}
        \includegraphics[width=\linewidth]{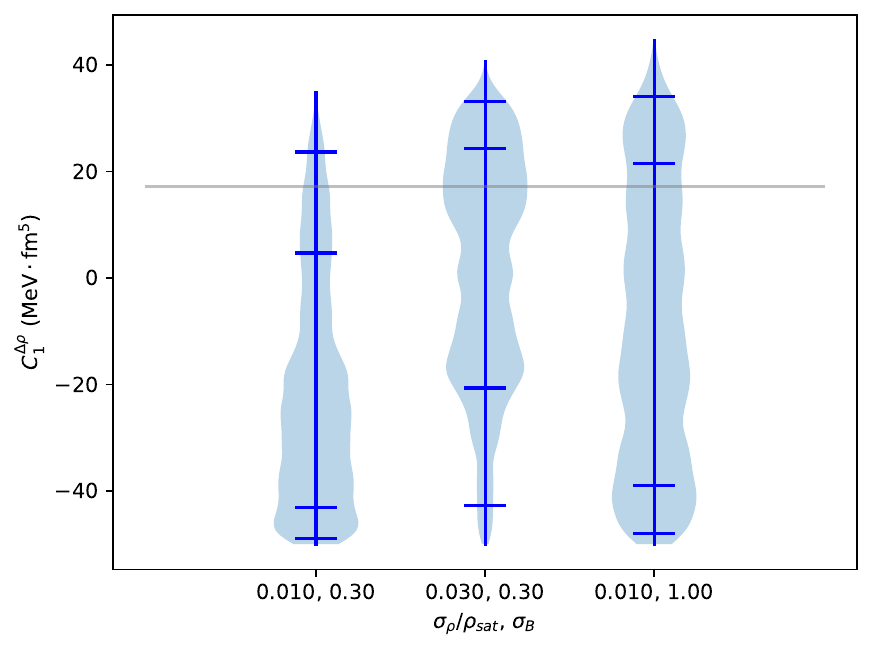}
    \end{subfigure}
    \caption{Violin plots of each Skyrme parameter for different $\sigma_{\rho}$ and $\sigma_B$. The EDF parameterization is SkM*. The unit of $\sigma_B$ is MeV. The short horizontal lines show the 68\% and 95\% credible bands. The long horizontal lines show the accurate values of SkM*.}
    \label{fig:violins-SkM*}
\end{figure*}
%%%%%%%%%%%%%%%%%%%%%%%%%%%%%%%%%%%%%%%%%%%%%%%%%%%

As the likelihood and prior are fixed in Secs.~\ref{ssec:likelihood} and \ref{ssec:prior}, respectively, the posterior is fully determined by Eq.~\eqref{eq:Baye}.
However, as $\vartheta$ is 12-dimensional (10 Skyrme parameters plus $\sigma_{\rho}'$ and $\sigma_B'$), one cannot obtain $Z$ by naively integrating $P(X|\vartheta) P(\vartheta)$ over the $\vartheta$ space.
Instead, we choose to sample $P(\vartheta|X)$ using the Markov chain Monte Carlo algorithm.
The major steps are:
\begin{enumerate}
    \item Start with an empty sample set $S=\{\}$.
    \item Add $\vartheta^{(0)}=(\theta^{(0)},\, \sigma_{\rho}^{\prime(0)},\,  \sigma_{B}^{\prime(0)})$ to $S$.\label{item:add-to-S}
    \item Propose $\vartheta^{(1/2)}=(\theta^{(1)},\,  \sigma_{\rho}^{\prime(0)},\,  \sigma_{B}^{\prime(0)})$. Here, $\theta^{(1)}=\theta^{(1)}+\delta\theta$, $\delta\theta\sim \mathcal{N}(0,s_{\theta})$, where $s_{\theta}$ is a metaparameter quantifying the sampling step size in the $\theta$ direction.\label{item:propose-theta}
    \item Compute the acceptance rate
    \begin{equation}
        \alpha = \max\left(1, \frac{P(X|\vartheta^{(1/2)})P(\vartheta^{(1/2)})}{P(X|\vartheta^{(0)})P(\vartheta^{(0)})}\right),
    \end{equation}
    and accept $\vartheta^{(1/2)}$ with probability $\alpha$.
    If accepted, go to the next step; otherwise, return to step~\ref{item:propose-theta}.
    \item Propose $\vartheta^{(1)}=(\theta^{(1)},\,  \sigma_{\rho}^{\prime(1)},\,  \sigma_{B}^{\prime(1)})$ with $\log\sigma_{\rho}^{\prime(1)}=\log\sigma_{\rho}^{\prime(0)}+\delta\eta$, $\log\sigma_{B}^{\prime(1)}=\log\sigma_{B}^{\prime(0)}+\delta\eta'$, $\delta\eta,\,\delta\eta'\sim \mathcal{N}(0,s_{\eta})$, where $s_{\eta}$ is a metaparameter quantifying the sampling step size in the $\log\sigma_{\rho}'$ and $\log\sigma_{B}'$ directions.\label{item:propose-sigma}
    \item Compute the acceptance rate
    \begin{equation}
        \alpha' = \max\left(1, \frac{P(X|\vartheta^{(1)})P(\vartheta^{(1)})}{P(X|\vartheta^{(1/2)})P(\vartheta^{(1/2)})}\right),
    \end{equation}
    and accept $\vartheta^{(1)}$ with probability $\alpha'$.
    If accepted, add $\vartheta^{(1)}$ to $S$; otherwise, return to step~\ref{item:propose-sigma}.
    \item Let $\vartheta^{(1)}$ to be the new $\vartheta^{(0)}$ and return to step~\ref{item:propose-theta}.
\end{enumerate}

In the above algorithm, we update $\theta$ and $\sigma_{\rho}'$, $\sigma_B'$ in separate steps following the Metropolis-within-Gibbs idea.
The main advantage is that during the updates of $\sigma_{\rho}'$ and $\sigma_B'$, $\theta$ is fixed, so HFBRAD is not executed again and again when the proposed $\sigma_{\rho}'$ and $\sigma_B'$ are rejected.
As the most time-consuming step of the algorithm is running HFBRAD, this strategy effectively reduces the running time of the code.

For the metaparameter $s_{\theta}$, we choose it adaptively such that $\sim20$\% of the proposed samples are accepted.
Once we find an optimal $s_{\theta}$, we fix it and start the sampling process.
For the metaparameter $s_{\eta}$, we choose $s_{\eta}=0.1$, so $\sigma_{\rho}'$ and $\sigma_B'$ update around 10\% at each sampling steps.
Hence, the sampling steps of  $\sigma_{\rho}'$ and $\sigma_B'$ become smaller in the low-$\sigma'$ region, allowing the algorithm to capture the drastic change of the likelihood function in the region.

Furthermore, we discovered that HFBRAD fails to converge for some values of $\theta$ allowed by the prior.
If the sampling chain starts in a region with a high density of non-convergent $\theta$, the sampling process drastically slows down due to the long waiting time before HFBRAD reports failure.
To avoid this type of region, we implement a genetic algorithm to choose the initial $\vartheta^{(0)}$.
The strategy is:
\begin{enumerate}
    \item Randomly pick many $\vartheta^{(0)}$ on the parameter space.
    \item If HFBRAD converges with $\theta^{(0)}$, use $P(X|\vartheta^{(0)})P(\vartheta^{(0)})$ as the fitness; otherwise set the fitness to 0.\label{item:fit}
    \item Pick the high-fitness $\vartheta^{(0)}$ and combine them into pairs to produce a new generation of $\vartheta^{(0)}$.\label{item:crossover}
    \item \label{item:mut}Introduce random mutations to the new generation and repeat steps~\ref{item:fit}--\ref{item:mut}.
\end{enumerate}
After several generations, the genetic algorithm successfully reduces the failure rate of HFBRAD to approximately zero.
Then, we choose around 10 $\vartheta^{(0)}$ from the final generation and use them as initial steps to start multiple sampling chains.
We discovered that the genetic algorithm also accelerates the sampling speed by reducing the burn-in time of each chain, during which the expectation value of $\theta$ drifts towards the accurate mean of the distribution.

Combining the above measures, we successfully get a sample set $S$ in which the density of samples is proportional to the magnitude of the posterior.
The expectation value of any observable is
\begin{equation}
    \langle O \rangle \equiv \int_{\vartheta} \, O(\vartheta) P(\vartheta|X)
    = \frac{1}{N_{\vartheta}} \sum_{\vartheta\in S} O(\vartheta),
\end{equation}
with $N_{\vartheta}$ being the total number of samples in $S$ and $O(\vartheta)$ being the value of $O$ predicted with $\vartheta$.
Similarly, one can define the median and credible bands of $O$.

%%%%%%%%%%%%%%%%%%%%%%%%%%%%%%%%%%%%%%%%%%%
\section{Results}
\label{sec:results}
%%%%%%%%%%%%%%%%%%%%%%%%%%%%%%%%%%%%%%%%%%%%%

%%%%%%%%%%%%%%%%%%%%%%%%%%%%%%%%%%%%%%%%%%%%%%%%%%%
\begin{figure}
    \centering
    \includegraphics[width=\linewidth]{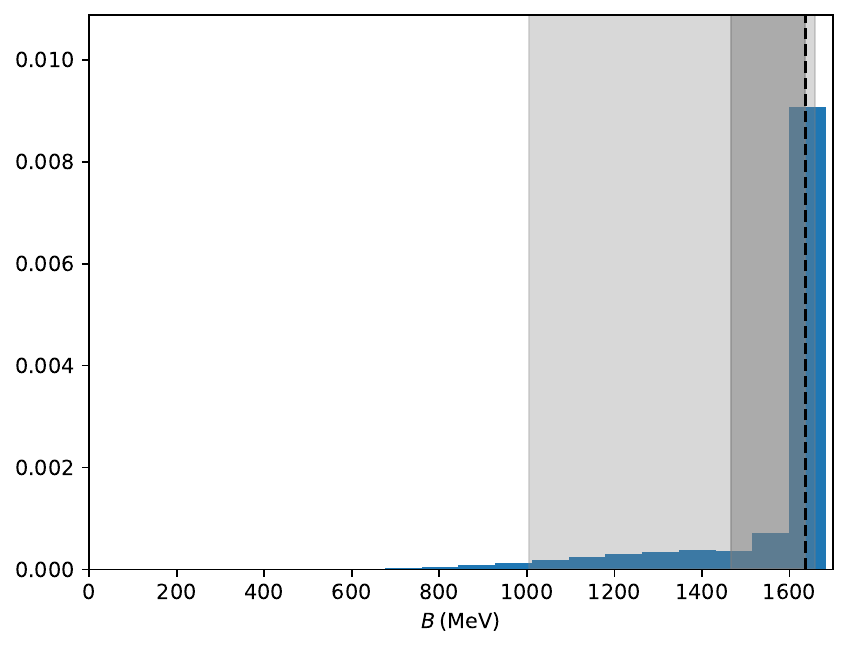}
    \caption{Predicted binding energy of $^{208}$Pb. The EDF parameterization is SkM*. The input noises are $\sigma_{\rho}=0.01\rho_{\text{sat}}$, $\sigma_B = 0.30\unit{MeV}$. The dashed line shows the accurate result. The shaded regions show the 68\% and 95\% credible bands, respectively.}
    \label{fig:Pb-B-SkMStar}
\end{figure}
%%%%%%%%%%%%%%%%%%%%%%%%%%%%%%%%%%%%%%%%%%%%%%%%%%%

%%%%%%%%%%%%%%%%%%%%%%%%%%%%%%%%%%%%%%%%%%%%%%%%%%%
\begin{figure*}
    \centering
    \begin{subfigure}{0.49\linewidth}
        \includegraphics[width=\linewidth]{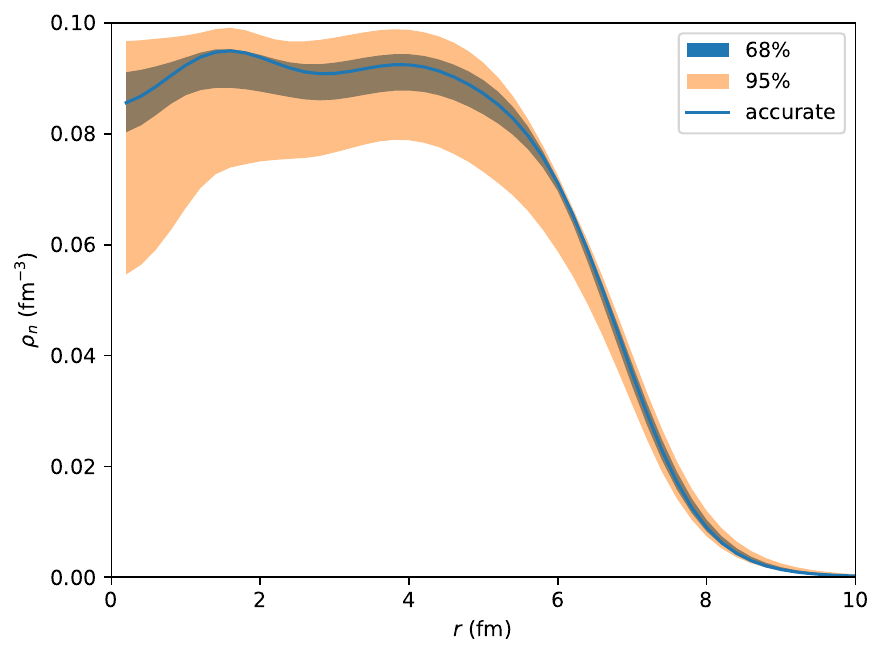}
        \caption{}
    \end{subfigure}
    \begin{subfigure}{0.49\linewidth}
        \includegraphics[width=\linewidth]{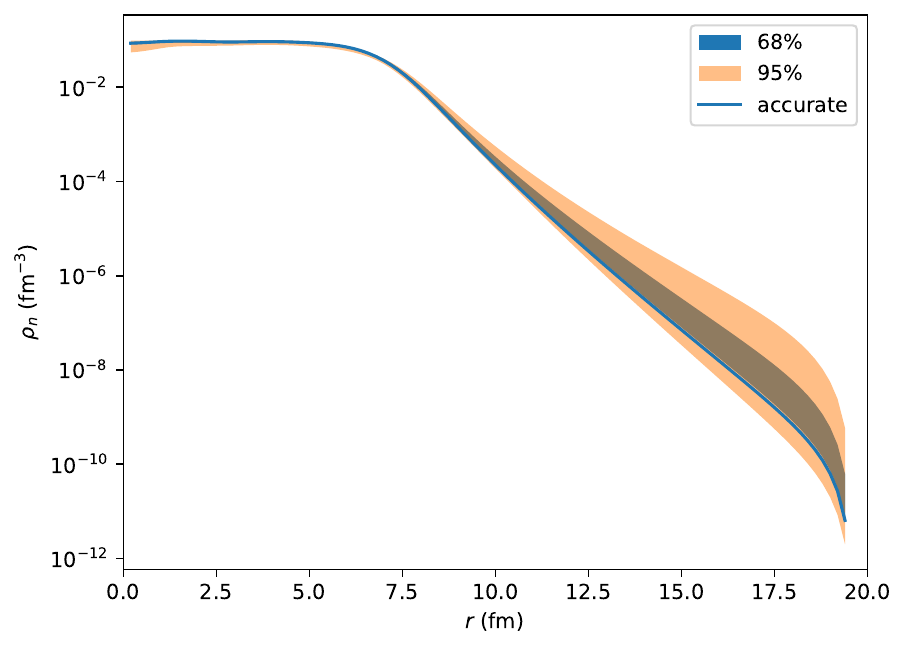}
        \caption{}
    \end{subfigure}
    \begin{subfigure}{0.49\linewidth}
        \includegraphics[width=\linewidth]{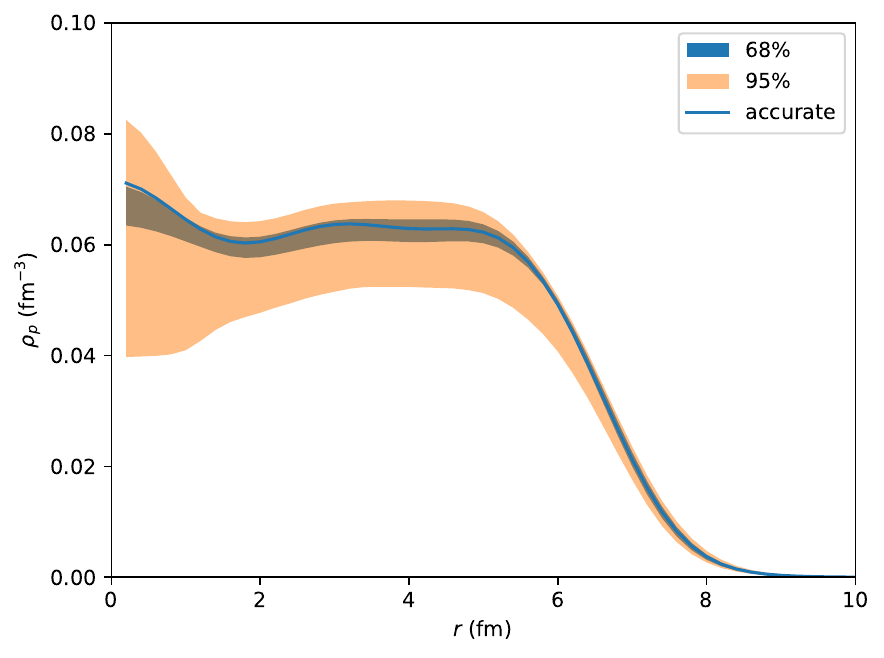}
        \caption{}
    \end{subfigure}
    \begin{subfigure}{0.49\linewidth}
        \includegraphics[width=\linewidth]{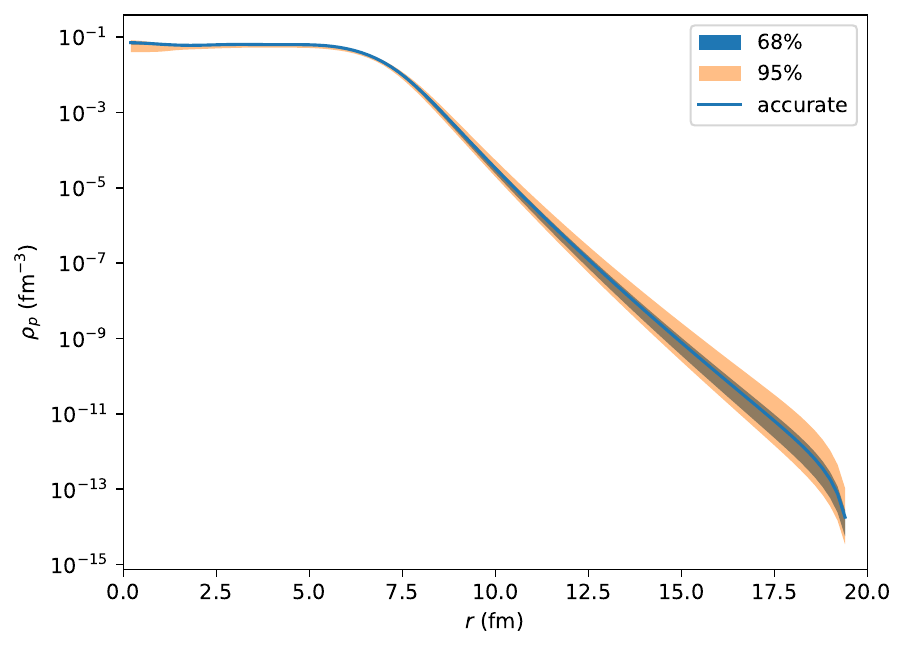}
        \caption{}
    \end{subfigure}
    \begin{subfigure}{0.49\linewidth}
        \includegraphics[width=\linewidth]{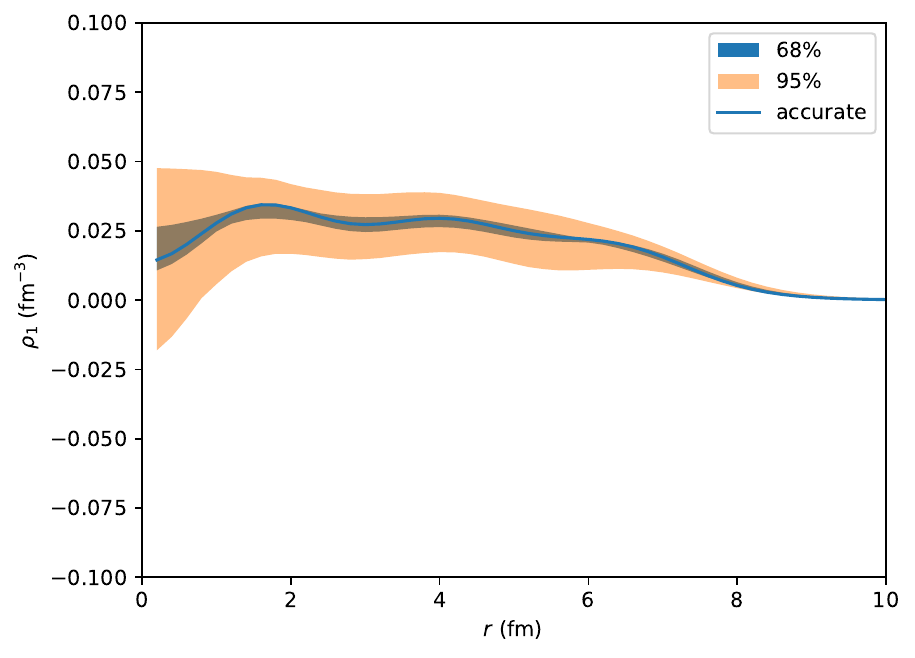}
        \caption{}
    \end{subfigure}
    \begin{subfigure}{0.49\linewidth}
        \includegraphics[width=\linewidth]{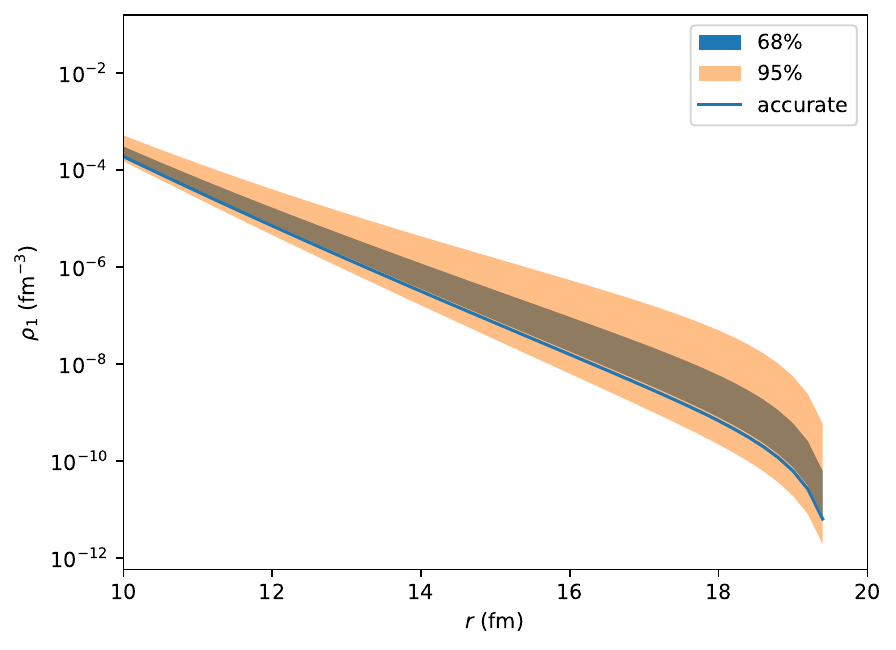}
        \caption{}
    \end{subfigure}
    \caption{Predicted 68\% and 95\% credible bands of neutron, proton, and isovector densities of $^{208}$Pb. The EDF parameterization is SkM*. The input noise is $\sigma_{\rho}=0.01\rho_{\text{sat}}$, $\sigma_B = 0.30\unit{MeV}$. The accurate SkM* results are also presented as a comparison.}
    \label{fig:Pb-rho-SkMStar}
\end{figure*}
%%%%%%%%%%%%%%%%%%%%%%%%%%%%%%%%%%%%%%%%%%%%%%%%%%%

%%%%%%%%%%%%%%%%%%%%%%%%%%%%%%%%%%%%%%%%%%%%%%%%%%%
\begin{figure}
    \centering
    \includegraphics[width=\linewidth]{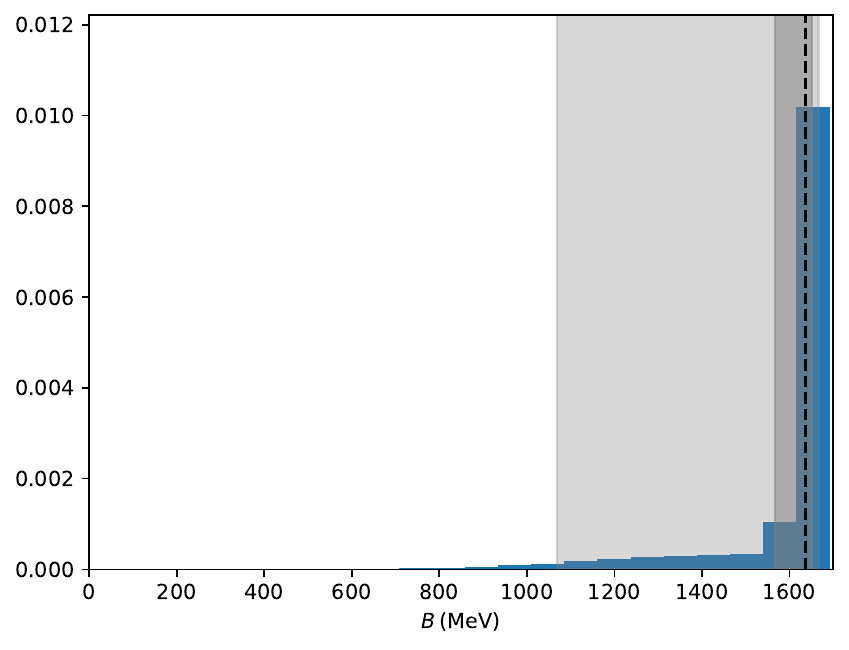}
    \caption{Predicted binding energy of $^{208}$Pb. The EDF parameterization is SLy4. The input noises are $\sigma_{\rho}=0.03\rho_{\text{sat}}$, $\sigma_B = 0.30\unit{MeV}$. The dashed line shows the accurate results. The shaded regions show the 68\% and 95\% credible bands, respectively.}
    \label{fig:Pb-B-SLy4}
\end{figure}
%%%%%%%%%%%%%%%%%%%%%%%%%%%%%%%%%%%%%%%%%%%%%%%%%%%

%%%%%%%%%%%%%%%%%%%%%%%%%%%%%%%%%%%%%%%%%%%%%%%%%%%
\begin{figure*}
    \centering
    \begin{subfigure}{0.49\linewidth}
        \includegraphics[width=\linewidth]{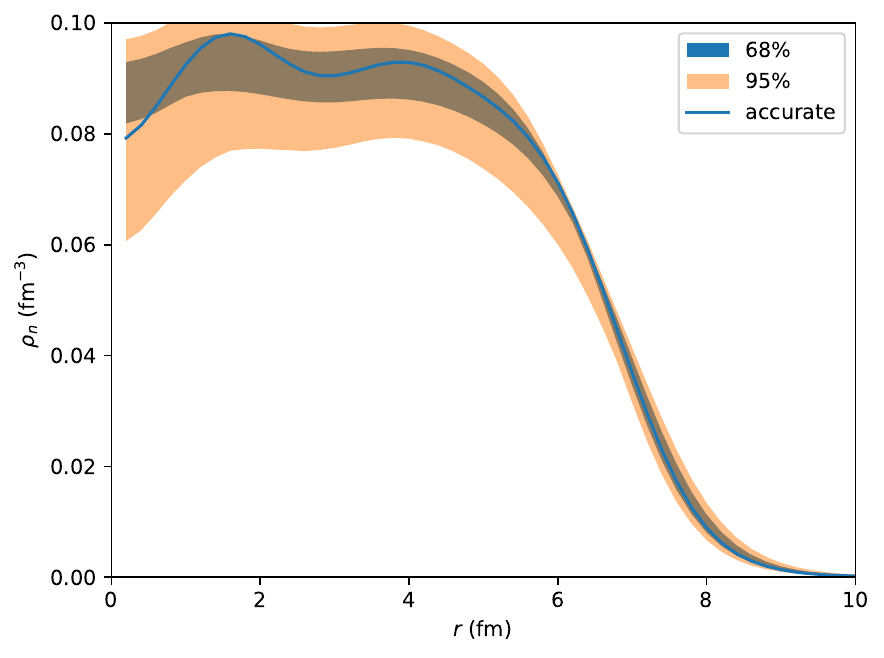}
        \caption{}
    \end{subfigure}
    \begin{subfigure}{0.49\linewidth}
        \includegraphics[width=\linewidth]{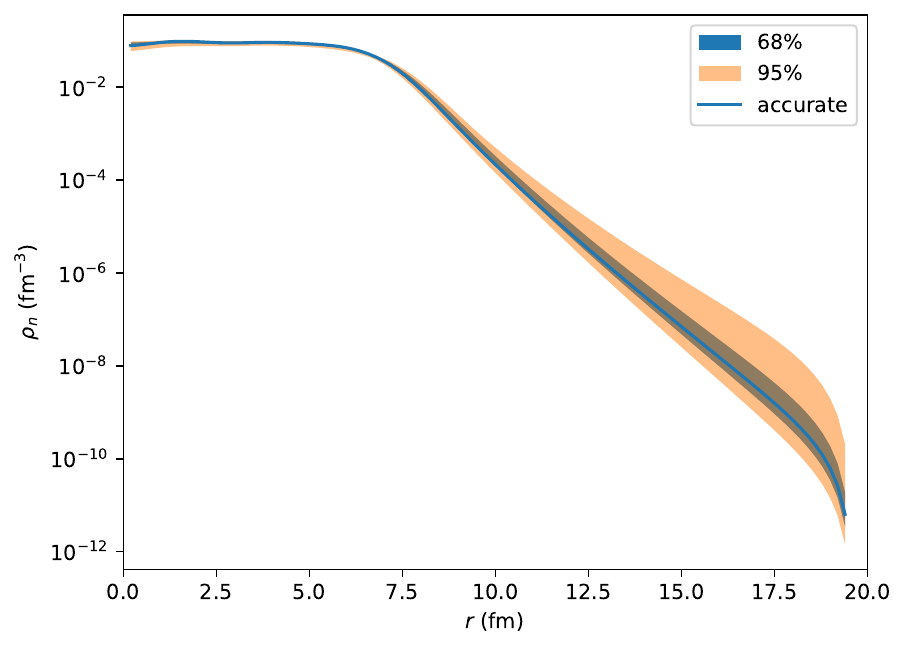}
        \caption{}
    \end{subfigure}
    \begin{subfigure}{0.49\linewidth}
        \includegraphics[width=\linewidth]{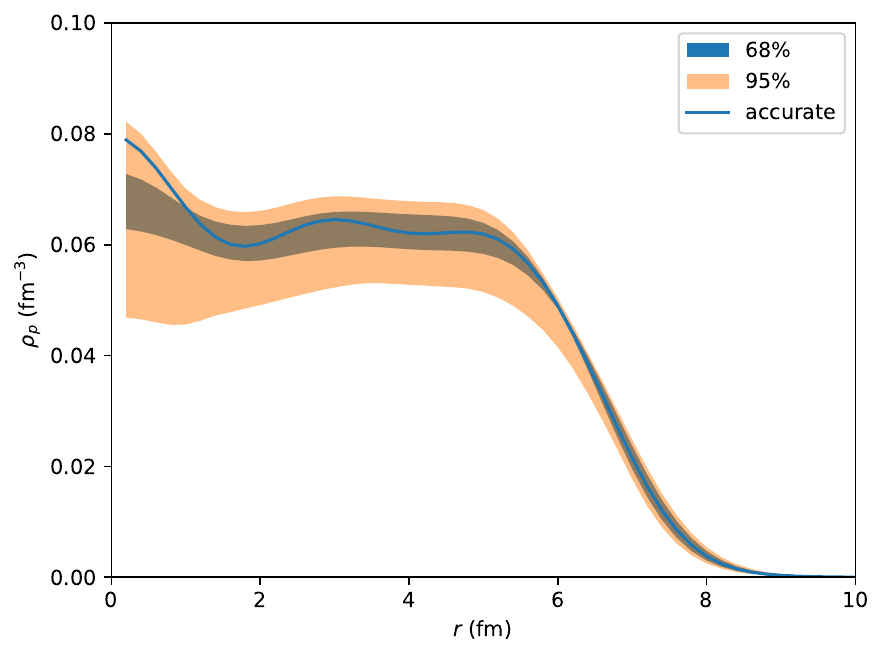}
        \caption{}
    \end{subfigure}
    \begin{subfigure}{0.49\linewidth}
        \includegraphics[width=\linewidth]{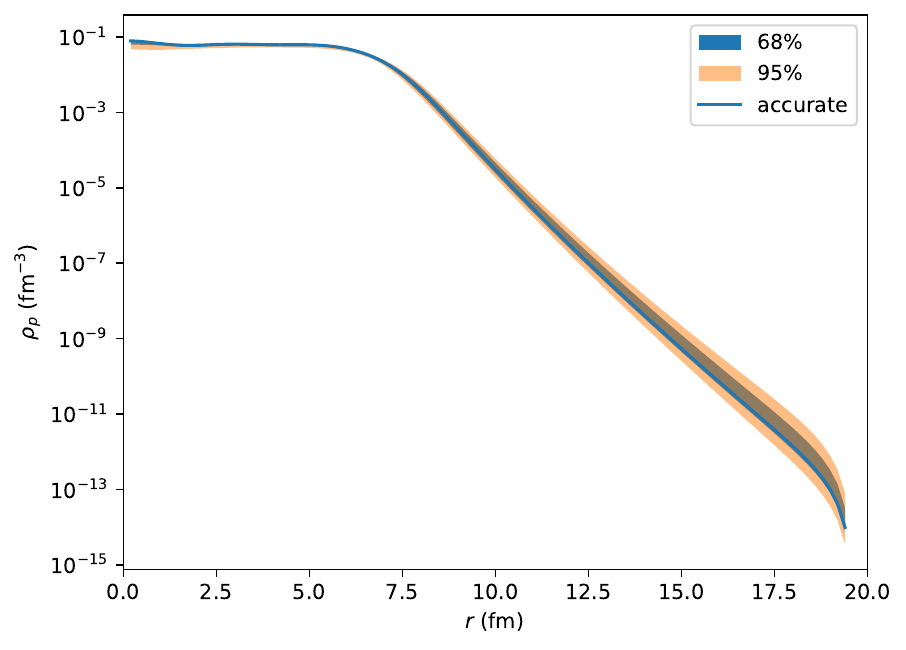}
        \caption{}
    \end{subfigure}
    \begin{subfigure}{0.49\linewidth}
        \includegraphics[width=\linewidth]{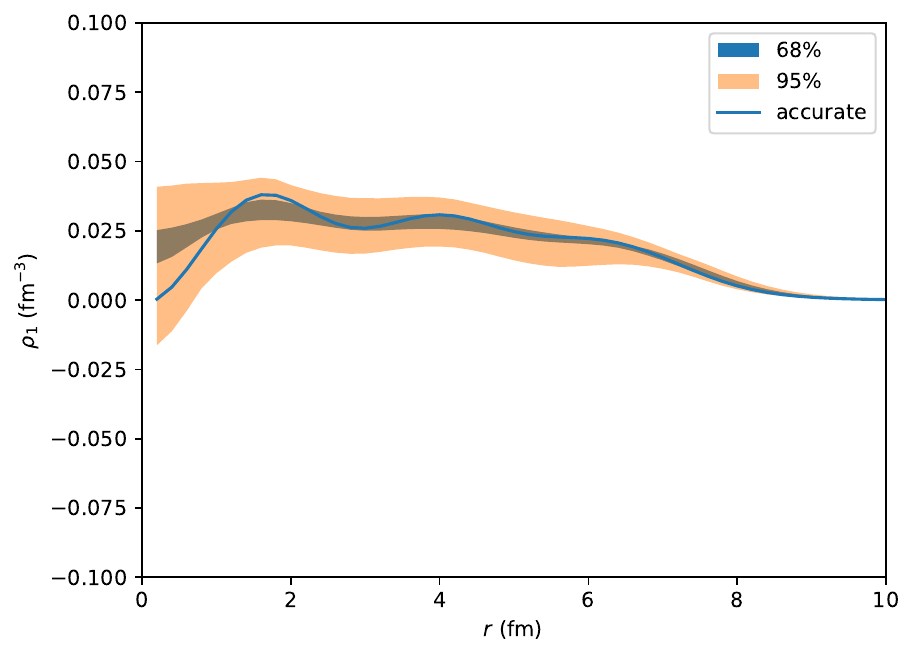}
        \caption{}
    \end{subfigure}
    \begin{subfigure}{0.49\linewidth}
        \includegraphics[width=\linewidth]{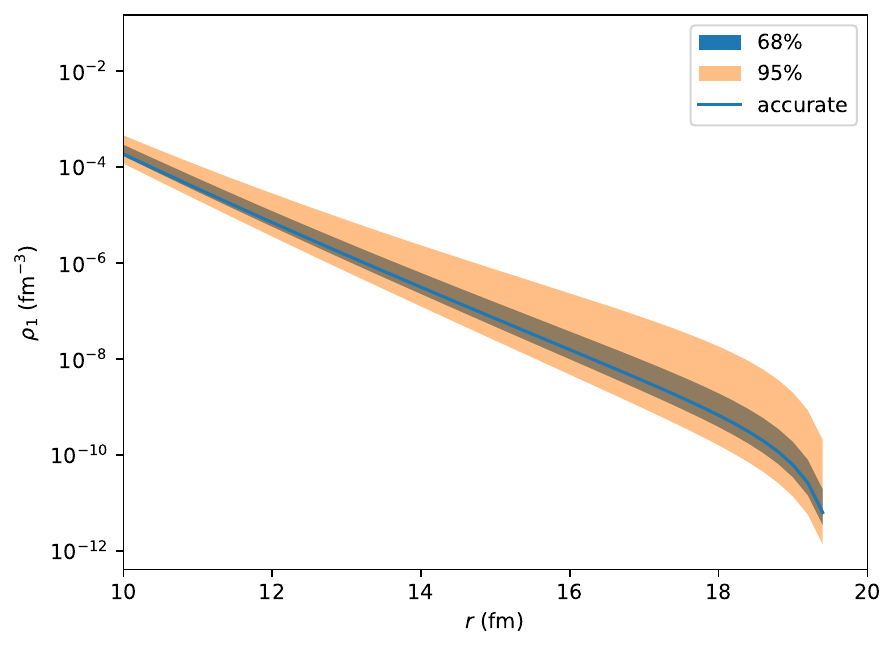}
        \caption{}
    \end{subfigure}
    \caption{Predicted 68\% and 95\% credible bands of neutron, proton, and isovector densities of $^{208}$Pb. The EDF parameterization is SLy4. The input noise is $\sigma_{\rho}=0.03\rho_{\text{sat}}$, $\sigma_B = 0.30\unit{MeV}$. The accurate SLy4 results are also presented as a comparison.}
    \label{fig:Pb-rho-SLy4}
\end{figure*}
%%%%%%%%%%%%%%%%%%%%%%%%%%%%%%%%%%%%%%%%%%%%%%%%%%%

In this section, we present the results of the Bayesian framework. 

%%%%%%%%%%%%%%%%%%%%%%%%%%%%%%%%%%%%%%%
\subsection{Input data}
\label{ssec:input}

In this work, we do not prepare the input data using ab initio methods like the no-core shell model.
This is because although ab initio results for $^{16}$O, $^{40}$Ca, and $^{48}$Ca exist (see Refs.~\cite{Roth:2007sv, Burrows:2017wqn, Burrows:2018ggt, Foy:2025yot}), the density profiles have not been systematically evaluated in a unified framework as far as we know.
Instead, we use HFBRAD to emulate ab initio input data.
In the baseline calculation, we use the SkM* parameterization of Skyrme EDF to prepare $\rho^{(N,Z)}_{n,p}$ and $B^{(N,Z)}$, and choose $\sigma_{\rho}=0.01\rho_{\text{sat}}$, $\sigma_B=0.3\unit{MeV}$.
We also study the SLy4 parameterization and change the value of $\sigma_{\rho}$ and $\sigma_B$ to compare with the baseline results.

%%%%%%%%%%%%%%%%%%%%%%%%%%%%%%%%%%%%%%%%%%%%%%%%%
\subsection{Inferred Skyrme parameters}
\label{ssec:parameters}

From the baseline input data, we obtain the 1-parameter distributions and 2-parameter correlations of $(\Theta,\,\sigma_{\rho}',\,\sigma_B')$ as a corner plot in Fig.~\ref{fig:corner}.
We vary the values $\sigma_{\rho}$ and $\sigma_B$ and make the 1-parameter distributions  of $\Theta$ as violin plots in Fig.~\ref{fig:violins-SkM*}.

\begin{table}
  \caption{Parameters of the phenomenological nuclear equation of state: accurate values (SkM*) and credible bands}
  \label{tab:cband}
  \begin{ruledtabular}
    \begin{tabular}{cccc}
      Name  &   accurate value  &   68\% band   &   95\% band\\
      \hline
      $\rho_{\text{sat}}\unit{(fm^{-3})}$  &   $0.16$    &   $0.153$--$0.163$    &     $0.151$--$0.171$\\
      $E_0\unit{(MeV)}$  &   $-15.77$    &   $-15.845$--$-15.644$     &    $-15.953$--$-15.554$\\
      $K_0\unit{(MeV)}$  &   $216.596$    &   $201.877$--$239.949$   &      $189.329$--$254.581$\\
      $J\unit{(MeV)}$  &   $30.033$    &   $29.515$--$36.026$     &    $26.367$--$37.540$\\
      $L\unit{(MeV)}$  &   $45.778$    &   $20.304$--$72.607$    &     $12.220$--$97.554$\\
    \end{tabular}
  \end{ruledtabular}
\end{table}

In these figures, $\rho_{\text{sat}}$ and $E_0$---the first two parameters from the phenomenological nuclear equation of state---show unimodal distributions.
For all $\sigma_{\rho}$ and $\sigma_B$, the accurate SkM* values of $\rho_{\text{sat}}$ and $E_0$ fall within the 68\% credible bands (see Fig.~\ref{fig:violins-SkM*} and Tab.~\ref{tab:cband}).
Thus, the density profiles and binding energies nicely constrain the properties of nuclear matter at equilibrium.

The remaining three parameters, $K_0$, $J$, and $L$, exhibit non-Gaussian or multi-modal distributions.
Nevertheless, in the baseline calculation, the accurate SkM* values of $K_0$, $J$, and $L$ fall within the 68\% credible bands (see Tab.~\ref{tab:cband}).
For all the $\sigma_{\rho}$ and $\sigma_B$ presented in Fig.~\ref{fig:violins-SkM*}, the accurate values fall within the 95\% credible bands, with the uncertainty of the parameters depending on both $\sigma_{\rho}$ and $\sigma_B$.
These results are particularly interesting because the input data in this work only includes the densities and binding energies of three nuclei, two of which have an equal number of neutrons and protons.
On the other hand, the original SkM* parameters are obtained by performing optimization on an extended amount of finite-nuclei observables; see Ref.~\cite{Bartel:1982ed}.
Even though the amount of input data is limited, the Bayesian frameworks predict the values of $K_0$, $J$, and $L$.
This shows that the density profiles of finite nuclei serve as a powerful tool to constrain off-equilibrium behavior of nuclear matter.

The Skyrme parameters related to non-uniform nuclear matter properties, $W_0$, $m_t^*$, and $C_t^{\Delta\rho}$, are harder to constrain.
Furthermore, their behavior exhibits strong isospin dependence.
While $W_0$, $m_0^*$, and $C_0^{\Delta\rho}$ are predicted within the 68\% credible bands, the uncertainties of $C_1^{\Delta\rho}$ and $m_1^*$ are significant compared with the prior in the current Bayesian framework.
One possible cause of these behaviors is that the parameters of non-uniform nuclear matter depend significantly on the off-diagonal part of the density matrices.
The information in this part vanishes in the density profiles.
In the future extension of this work, one may need to introduce more ab initio results to fully constrain the parameters.

As shown in Fig.~\ref{fig:corner}, the uncertainty of EDF ($\sigma_{\rho}'$ and $\sigma_B'$) is significantly smaller than the uncertainty of input data ($\sigma_{\rho}$ and $\sigma_B$) and shows long tails in the small-value region of the $\log$ plot.
This behavior occurs because this work uses HFBRAD to emulate the input data.
Hence, on the parameter space, there is a $\theta_0$ such that $\rho^{(N,Z)}_{q} - \rho^{(N,Z)}_q(\theta_0)=0$, and the likelihood shows large peak around $\vartheta=(\theta_0, 0, 0)$.
In the log scale we are using, the $(\sigma'_{\rho}, \sigma_B')=(0,0)$ point is at $(-\infty,-\infty)$.
Hence, the sampling process shows a continuous drift towards the $-\infty$ direction.
If we have real ab initio data in the future, the distribution may exhibit richer structure and allow us to investigate the boundary of the capacity of EDF.

%%%%%%%%%%%%%%%%%%%%%%%%%%%%%%%%%%%%%%%%%%%
\subsection{Prediction of $^{208}$Pb properties}
\label{ssec:predictions}

In the previous subsection, we use density profiles and binding energies to infer the probability distribution of Skyrme parameters.
From the perspective of an optimization task, the outcome is expected:
As we only use information from the diagonal part of the density matrices, the obtained distribution is multi-modal;
If we include more observables related to the off-diagonal part, as in the original work of SkM*, all parameters will eventually become well-constrained.
The crucial question, however, is not merely whether the Bayesian framework can reproduce the accurate values of the parameters, but whether the resulting parameter distribution can provide reliable predictions for observables \textit{not included in the input data}.
This predictive capability is a more stringent and non-trivial test of the framework.

To examine the predictive capability, we use the Skyrme parameters obtained to predict the behavior of $^{208}$Pb, another double-magic nucleus that is not included in the input data set.
We use HFBRAD to compute the density profiles and binding energy for each $\theta\in S$.
The distribution and credible bands of the binding energy are presented in Fig.~\ref{fig:Pb-B-SkMStar}.
The credible bands of the density profiles are presented in Fig.~\ref{fig:Pb-rho-SkMStar}.

Fig.~\ref{fig:Pb-B-SkMStar} shows a highly concentrated distribution of the binding energy with a long tail in the low-energy direction.
The accurate value of SkM* is $1636.404\unit{MeV}$.
The median is $1617.260\unit{MeV}$.
The 68\% credible band is $1466.340$--$1636.037\unit{MeV}$.
The 95\% credible band is $1005.929$--$1658.901\unit{MeV}$.
The proposed Bayesian framework shows evident predictive capability for the binding energy.

In the same way, Fig.~\ref{fig:Pb-rho-SkMStar} shows that in the major region of the coordinate space, the accurate density profiles fall within the 68\% credible bands.
For the whole space, they fall within the 95\% credible bands.
The proposed Bayesian framework shows evident predictive capability for the density profiles.

Interestingly, we notice that the density profiles in the surface region of $^{208}$Pb remain well-constrained (see Fig.~\ref{fig:Pb-rho-SkMStar}(a), (c) for the $\rho_{n,p}$ in the linear scale and Fig.~\ref{fig:Pb-rho-SkMStar}(b), (d) for $\rho_{n,p}$ in the log scale).
The well-behaved tail of the distribution originates from the likelihood defined in Eq.~\eqref{eq:likelihood}---when computing the likelihood, we weight $|\rho_t^{(N, Z)}(r_i)-\rho_t^{(N, Z)}(r_i;\theta)|$ with $\rho_t^{(N, Z)}(r_i) r_i^2$, which reflects the large space volume that the surface region occupies.
This behavior is advantageous for predicting crucial experimental observables that depend on the densities in the surface region, e.g., the neutron skin thickness, which quantifies the difference between the neutron and proton radii, and the dipole polarizability, which quantifies the response of nuclei to external electric dipole fields.
The observables are particularly sensitive with the isovector density $\rho_1(r)=\rho_n(r)-\rho_p(r)$ near the surface.
We also plot the credible bands of $\rho_1(r)$ in the center and surface regions in Fig.~\ref{fig:Pb-rho-SkMStar}(d), (e), respectively.
We confirm that the accurate values of $\rho_1$ are well-constrained by the Bayesian framework.

To avoid the overfitting of the metaparameters to the validation set during the model-development stage, we use the Bayesian framework to predict the behavior of $^{208}$Pb under a different EDF parameterization---SLy4. Furthermore, we change the input uncertainties to $\sigma_{\rho} = 0.03\rho_{\text{sat}}$ and $\sigma_B = 0.30\unit{MeV}$.
The predicted binding energy is shown in Fig.~\ref{fig:Pb-B-SLy4}.
The predicted density profiles are in Fig.~\ref{fig:Pb-rho-SLy4}.
In these figures, the accurate results of the binding energy and density profiles are still within the credible bands predicted. 
The qualitative behavior of the bands is the same as the baseline case in both the center and surface regions.
This verifies that the predictive capability of the Bayesian framework is robust over different setups.

%%%%%%%%%%%%%%%%%%%%%%%%%%%%%%%%%%%%%%%%%%%%
\section{Conclusions}
\label{sec:concl}
%%%%%%%%%%%%%%%%%%%%%%%%%%%%%%%%%%%%%%%%%%%%
In this work, we use the density profiles and binding energies of finite nuclei to constrain parameters of the Skyrme EDF.
We establish a Bayesian analysis framework in which we
1) transform the Skyrme parameters into a 10-dimensional physically meaningful representation and construct a prior based on it, and
2) construct the likelihood based on the difference between input data and HFBRAD calculation results, with the uncertainties of the input and HFBRAD calculation taken into account independently.
We show that:
\begin{itemize}
    \item With density profiles, the Bayesian framework effectively learns all the parameters relating to uniform nuclear matter properties.
    The properties of non-uniform nuclear matter remain partially constrained and require more input.
    \item Trained on data of light nuclei, the Bayesian framework predicts the densities and binding energy of $^{208}$Pb.
    In particular, it yields well-constrained behavior in the surface region of the nuclei, which is important for predicting experimental observables.
\end{itemize}
In nuclear many-body physics, ab initio methods provide accurate density profiles of light nuclei.
On the other hand, the EDF method is among the most effective tools for heavy nuclei.
By constraining EDF from density profiles of light nuclei, the Bayesian framework established here serves as a unique tool to connect the two fields.

Possible future extensions of this work include 1) incorporating different EDFs (e.g., Skyrme, Gogny, and covariant EDFs) in the same Bayesian framework,
2) incorporating real ab initio data as input,
and 3) generalizing the kernel function of the likelihood, systematically considering the correlation of nearby points of the density profiles.

%%%%%%%%%%%%%%%%%%%%%%%%%%%%%%%%%
\acknowledgments
%%%%%%%%%%%%%%%%%%%%%%%%%%%%%%%%%%
We would like to thank the members of the QMS project, in particular, E.~Hiyama, T.~Nakatsukasa, M.~Oka, K.~Sekizawa, N.~Shimizu, and Y.~Tsunoda, for valuable discussion at the early stage of the work. This work was supported by
JSPS KAKENHI (Grant Nos. JP23K25864, JP25K07322, JP25K24695, JP25H01269, JP26H02035), JST ERATO (Grant No. JPMJER2304). This work also used computational resources of the supercomputer SQUID, from the Large-Scale Computer System operated by D3 Center, provided by RCNP of the University of Osaka.

%%%%%%%%%%%%%%%%%%%%%%
\appendix
%%%%%%%%%%%%%%%%%%%%%%
\section{Mapping between the $\theta$ and $\Theta$}
\label{app:trans}

In uniform nuclear matter, $\bm{J}_t=0$.
From Eq.~\eqref{eq:Skyrme}, one obtain the energy density as
\begin{equation}
    E = \frac{\hbar^{2}\tau_{0}}{2m} + \mathcal{E}[\rho] = \frac{\hbar^{2}\tau_{0}}{2m}+\sum_{t}\left[(C_{t0}^{\rho}+C_{tD}^{\rho}\rho_{0}^{\gamma})\rho_{t}^{2}+C_{t}^{\tau}\rho_{t}\tau_{t}\right].
    \label{eq:E-uniform}
\end{equation}
Here, we ignore the Coulomb sector.

According to the standard argument based on Fermi momentum, 
\begin{equation}
    \tau_{n,p}=C_{k}'\rho_{n,p}^{5/3},
\end{equation}
so
\begin{equation}
   \tau_{0}=\tau_{n}+\tau_{p}=C_{k}\rho_{0}^{5/3}+\frac{5}{9}C_{k}\rho_{0}^{5/3}\beta^{2}+O(\beta^{3}),
    \label{eq:tau0-uniform}
\end{equation}
\begin{equation}
    \tau_{1}=\tau_{n}-\tau_{p}=\frac{5}{3}C_{k}\rho_{0}^{5/3}\beta+O(\beta^{3}).
    \label{eq:tau1-uniform}
\end{equation}
Here, $C_k=3/5 (3\pi^2/2)^{2/3}$ and $C_k'=(3/5) (3\pi^2)^{2/3}$. 
As introduced in the main text, $\beta=(\rho_n-\rho_p)/\rho_0$.

With Eqs.\eqref{eq:E-uniform}--\eqref{eq:tau1-uniform}, the energy per particle is
\begin{align}
    e(\rho_{0},\beta)=e(\rho_{0})+S_{2}(\rho_{0})\beta^{2}+O(\beta^{4}),
    \label{eq:e-S2}
\end{align}
\begin{equation}
    e(\rho_{0})=\frac{\hbar^{2}}{2m}C_{k}\rho_{0}^{2/3}+C_{00}^{\rho}\rho_{0}+C_{0D}^{\rho}\rho_{0}^{\gamma+1}+C_{k}C_{0}^{\tau}\rho_{0}^{5/3},
\end{equation}
\begin{align}
    &S_{2}(\rho_{0})=\frac{5\hbar^{2}}{18m}C_{k}\rho_{0}^{2/3}+\frac{5}{9}C_{k}C_{0}^{\tau}\rho_{0}^{5/3}\eqnl+C_{10}^{\rho}\rho_{0}+C_{1D}^{\rho}\rho_{0}^{\gamma+1}+\frac{5}{3}C_{k}C_{1}^{\tau}\rho_{0}^{5/3}.
\end{align}

The nuclear saturation density is determined by $\partial e(\rho_0)/\partial\rho_0 = 0$, this yields the following equation,
\begin{equation}
    \frac{\hbar^{2}}{3m}C_{k}\rho_{\text{sat}}^{2/3}+C_{00}^{\rho}\rho_{\text{sat}}+C_{0D}^{\rho}(\gamma+1)\rho_{\text{sat}}^{\gamma+1}+\frac{5}{3}C_{k}C_{\text{sat}}^{\tau}\rho_{\text{sat}}^{5/3}=0.
    \label{eq:rhosat}
\end{equation}

Comparing Eq.~\eqref{eq:e-S2} with Eq.~\eqref{eq:eos}, one confirms that
\begin{equation}
    E_{0}=e(\rho_{\text{sat}})=\frac{\hbar^{2}}{2m}C_{k}\rho_{\text{sat}}^{2/3}+C_{00}^{\rho}\rho_{\text{sat}}+C_{0D}^{\rho}\rho_{\text{sat}}^{\gamma+1}+C_{k}C_{0}^{\tau}\rho_{\text{sat}}^{5/3},
    \label{eq:E0}
\end{equation}
\begin{align}
    &K_{0}=9\rho_{\text{sat}}^{2}\left.\frac{\partial^{2}e(\rho)}{\partial\rho^{2}}\right|_{\rho_{0}=\rho_{\text{sat}}}=-\frac{\hbar^{2}}{m}C_{k}\rho_{\text{sat}}^{2/3}\eqnl+C_{0D}^{\rho}9(\gamma+1)\gamma\rho_{\text{sat}}^{\gamma+1}+10C_{0}^{\tau}C_{k}\rho_{\text{sat}}^{5/3},
    \label{eq:K0}
\end{align}
\begin{align}
    &J=S_{2}(\rho_{\text{sat}})=\frac{5\hbar^{2}}{18m}C_{k}\rho_{\text{sat}}^{2/3}+\frac{5}{9}C_{k}C_{0}^{\tau}\rho_{\text{sat}}^{5/3}\eqnl+C_{10}^{\rho}\rho_{\text{sat}}+C_{1D}^{\rho}\rho_{\text{sat}}^{\gamma+1}+\frac{5}{3}C_{k}C_{1}^{\tau}\rho_{\text{sat}}^{5/3},
    \label{eq:J}
\end{align}
\begin{align}
    &L=3\rho_{\text{sat}}\left.\frac{\partial S_{2}(\rho_{0})}{\partial\rho_{0}}\right|_{\rho_{0}=\rho_{\text{sat}}}=\frac{5\hbar^{2}}{9m}C_{k}\rho_{\text{sat}}^{2/3}+\frac{25}{9}C_{k}C_{0}^{\tau}\rho_{\text{sat}}^{5/3}\eqnl
    +3C_{10}^{\rho}\rho_{\text{sat}}+3C_{1D}^{\rho}(\gamma+1)\rho_{\text{sat}}^{\gamma+1}+\frac{25}{3}C_{k}C_{1}^{\tau}\rho_{\text{sat}}^{5/3}.
    \label{eq:L}
\end{align}

The definition of effective mass (Eqs.~\eqref{eq:effmass-m0}, \eqref{eq:effmass-m1}) and the Skyrme EDF (Eq.~\eqref{eq:Skyrme}) show
\begin{equation}
    \frac{m^{*}_{0}}{m}=\frac{1}{1+\frac{2m}{\hbar^{2}}C^{\tau}_{0}\rho_{\text{sat}}},
    \label{eq:effmass-m0-simp}
\end{equation}
and 
\begin{equation}
    \frac{m^{*}_{1}}{m}=\frac{1}{1+\frac{2m}{\hbar^{2}}\left(C^{\tau}_{0}-C^{\tau}_{1}\right)\rho_{\text{sat}}}.
    \label{eq:effmass-m1-simp}
\end{equation}

Eq.~\eqref{eq:interaction-so} and the $\rho_t \nabla\cdot\bm{J}_t$ term in the Skyrme EDF describe the same interaction.
By comparing these terms, one notices that
\begin{equation}
    C^{\nabla J}_{0}=-\frac{3}{4}W_{0},\,C^{\nabla J}_{1}=-\frac{1}{4}W_{0}.
    \label{eq:W0}
\end{equation}

Eqs.~\eqref{eq:rhosat}--\eqref{eq:W0} constitute a one-to-one mapping $f$ between $\theta = (C^{\rho}_{00},\, C^{\rho}_{0D},\, C_0^{\Delta\rho},\, C_0^{\tau},\, C_0^{\nabla J},\, C^{\rho}_{10},\, C^{\rho}_{1D},\, C_1^{\Delta\rho},\, C_1^{\tau},\, \gamma)$ and $\Theta = (\rho_{\text{sat}},\, E_0,\, K_0,\, J,\, L,\, m_0^*,\, m_1^*,\, W_0,\, C_0^{\Delta\rho},\, C_1^{\Delta\rho})$.
One can introduce the forward and backward mappings as below.

First, we start from $\theta$. 
One can numerically verify that in the physically meaningful regime around $\rho_{\text{sat}}\sim0.16\unit{fm^{-3}}$, Eq.~\eqref{eq:rhosat} has a unique solution.
With $\rho_{\text{sat}}$ known, Eqs.~\eqref{eq:E0}--\eqref{eq:L} determine $E_0$, $K_0$, $J$, and $L$, respectively;
Eqs.~\eqref{eq:effmass-m0-simp} and \eqref{eq:effmass-m1-simp} determine $m_0^*$ and $m_1^*$, respectively;
Eq.~\eqref{eq:W0} determines $W_0$;
and $C_0^{\Delta\rho}$, $C_1^{\Delta\rho}$ do not change in the mapping.
Thus, $\Theta$ is fully determined.

Then, we start from $\Theta$.
We use Eqs.~\eqref{eq:effmass-m0-simp} and \eqref{eq:effmass-m1-simp} to determine
\begin{equation}
    C^{\tau}_{0}=\frac{\hbar^{2}}{2\rho_{0}}\left(\frac{1}{m^{*}_{0}}-\frac{1}{m}\right),
\end{equation}
\begin{equation}
    C^{\tau}_{1}=\frac{\hbar^{2}}{2\rho_{0}}\left(\frac{1}{m^{*}_{0}}-\frac{1}{m^{*}_{1}}\right).
\end{equation}
Then, we combine Eqs.~\eqref{eq:rhosat}--\eqref{eq:K0} to eliminate $C_{00}^{\rho}$ and  $C_{0D}^{\rho}$, and determine
\begin{equation}
    \gamma=\frac{1}{9}\frac{K_{0}+\frac{\hbar^{2}}{m}C_{k}\rho_{\text{sat}}^{2/3}-10C_{k}C_{0}^{\tau}\rho_{\text{sat}}^{5/3}}{\frac{\hbar^{2}}{6m}C_{k}\rho_{\text{sat}}^{2/3}-\frac{2}{3}C_{k}C_{0}^{\tau}\rho_{\text{sat}}^{5/3}-E_{0}}-1.
\end{equation}
With $\gamma$ known, the remaining equations of Eqs.~\eqref{eq:rhosat}--\eqref{eq:L} is a linear system of $C_{00}^{\rho}$,  $C_{0D}^{\rho}$, $C^{\rho}_{1D}$, and $C^{\rho}_{1D}$.
We obtain
\begin{equation}
    C_{0D}^{\rho}=\frac{1}{\gamma\rho_{\text{sat}}^{\gamma+1}}\biggl(\frac{\hbar^{2}}{6m}C_{k}\rho_{\text{sat}}^{2/3}-\frac{2}{3}C_{k}C_{0}^{\tau}\rho_{\text{sat}}^{5/3}-E_{0}\biggr),
\end{equation}
\begin{equation}
    C_{00}^{\rho}=-\frac{1}{\rho_{\text{sat}}}\biggl(\frac{\hbar^{2}}{3m}C_{k}\rho_{\text{sat}}^{2/3}+\frac{5}{3}C_{k}C_{0}^{\tau}\rho_{\text{sat}}^{5/3}+C_{0D}^{\rho}(\gamma+1)\rho_{\text{sat}}^{\gamma+1}\biggr),
\end{equation}
\begin{align}
    &C_{1D}^{\rho}=\frac{1}{\gamma\rho_{\text{sat}}^{\gamma+1}}\biggl(\frac{1}{3}L-J+\frac{5\hbar^{2}}{54m}C_{k}\rho_{\text{sat}}^{2/3}\eqnl
    -\frac{10}{27}C_{k}C_{0}^{\tau}\rho_{\text{sat}}^{5/3}-\frac{10}{9}C_{k}C_{1}^{\tau}\rho_{\text{sat}}^{5/3}\biggr),
\end{align}
\begin{align}
    &C_{10}^{\rho}=\frac{1}{\rho_{\text{sat}}}\biggl(J-\frac{5\hbar^{2}}{18m}C_{k}\rho_{\text{sat}}^{2/3}-\frac{5}{9}C_{k}C_{0}^{\tau}\rho_{\text{sat}}^{5/3}\eqnl-\frac{5}{3}C_{k}C_{1}^{\tau}\rho_{\text{sat}}^{5/3}-C_{1D}^{\rho}\rho_{\text{sat}}^{\gamma+1}\biggr).
\end{align}
Finally, we use Eq.~\eqref{eq:W0} to determines $C^{\nabla J}_{t}$.
$C_0^{\Delta\rho}$, $C_1^{\Delta\rho}$ do not change in the mapping.
Thus, $\theta$ is completely determined.

This concludes the construction of the one-to-one mapping.    

%%%%%%%%%%%%%%%%%%%%%%%%%%%%%%%%%%%%%%%%%%%%%%%%
\bibliography{refs.bib}

@article{Bender:2003jk,
    author = "Bender, Michael and Heenen, Paul-Henri and Reinhard, Paul-Gerhard",
    title = "{Self-consistent mean-field models for nuclear structure}",
    doi = "10.1103/RevModPhys.75.121",
    journal = "Rev. Mod. Phys.",
    volume = "75",
    pages = "121--180",
    year = "2003"
}

@article{Nakatsukasa:2016nyc,
    author = "Nakatsukasa, Takashi and Matsuyanagi, Kenichi and Matsuo, Masayuki and Yabana, Kazuhiro",
    title = "{Time-dependent density-functional description of nuclear dynamics}",
    eprint = "1606.04717",
    archivePrefix = "arXiv",
    primaryClass = "nucl-th",
    doi = "10.1103/RevModPhys.88.045004",
    journal = "Rev. Mod. Phys.",
    volume = "88",
    number = "4",
    pages = "045004",
    year = "2016"
}

@article{Colo:2020vik,
    author = "Col{\`o}, G.",
    title = "{Nuclear density functional theory}",
    doi = "10.1080/23746149.2020.1740061",
    journal = "Adv. Phys. X",
    volume = "5",
    number = "1",
    pages = "Article: 1740061",
    year = "2020"
}

@article{Yang:2019fvs,
    author = "Yang, Junjie and Piekarewicz, J.",
    title = "{Covariant Density Functional Theory in Nuclear Physics and Astrophysics}",
    eprint = "1912.11112",
    archivePrefix = "arXiv",
    primaryClass = "nucl-th",
    doi = "10.1146/annurev-nucl-101918-023608",
    journal = "Ann. Rev. Nucl. Part. Sci.",
    volume = "70",
    pages = "21--41",
    year = "2020"
}

@article{Nakatsukasa:2025axc,
    author = "Nakatsukasa, Takashi and Yu, Chengpeng",
    title = "{Energy density functional approaches to inhomogeneous superfluid neutron-star matter}",
    doi = "10.22323/1.465.0170",
    journal = "PoS",
    volume = "QNP2024",
    pages = "170",
    year = "2025"
}

@article{Yu:2025hmc,
    author = "Yu, Chengpeng and Nakatsukasa, Takashi",
    title = "{Fermi operator expansion for the Hartree-Fock-Bogoliubov theory}",
    eprint = "2504.04735",
    archivePrefix = "arXiv",
    primaryClass = "nucl-th",
    doi = "10.1103/y65c-svnp",
    journal = "Phys. Rev. C",
    volume = "112",
    number = "1",
    pages = "015804",
    year = "2025"
}

@article{Vautherin:1971aw,
    author = "Vautherin, D. and Brink, D. M.",
    title = "{Hartree-Fock calculations with Skyrme's interaction. 1. Spherical nuclei}",
    doi = "10.1103/PhysRevC.5.626",
    journal = "Phys. Rev. C",
    volume = "5",
    pages = "626--647",
    year = "1972"
}

@article{Vautherin:1973zz,
    author = "Vautherin, D.",
    title = "{Hartree-Fock Calculations with Skyrme's Interaction. 2. Axially Deformed Nuclei}",
    doi = "10.1103/PhysRevC.7.296",
    journal = "Phys. Rev. C",
    volume = "7",
    pages = "296--316",
    year = "1973"
}

@article{Bartel:1982ed,
    author = "Bartel, J. and Quentin, P. and Brack, M. and Guet, C. and Hakansson, H. -B.",
    title = "{Towards a better parametrisation of Skyrme-like effective forces: A Critical study of the SkM force}",
    doi = "10.1016/0375-9474(82)90403-1",
    journal = "Nucl. Phys. A",
    volume = "386",
    pages = "79--100",
    year = "1982"
}

@article{Dzhioev:2025wey,
    author = "Dzhioev, Alan A. and Antonenko, N. V.",
    title = "{Isoenergetic description of induced fission pathways within energy-density functional theory}",
    eprint = "2509.22006",
    archivePrefix = "arXiv",
    primaryClass = "nucl-th",
    doi = "10.1103/w2rd-557s",
    journal = "Phys. Rev. C",
    volume = "112",
    number = "3",
    pages = "034333",
    year = "2025"
}

@article{An:2025twk,
    author = "An, Rong and Sun, Shuai and Jiang, Xiang and Tang, Na and Cao, Li-Gang and Zhang, Feng-Shou",
    title = "{Shell effects in nuclear charge radii based on Skyrme density functionals}",
    eprint = "2506.20414",
    archivePrefix = "arXiv",
    primaryClass = "nucl-th",
    doi = "10.1103/nyn5-69s3",
    journal = "Phys. Rev. C",
    volume = "112",
    number = "2",
    pages = "024303",
    year = "2025"
}

@article{Chabanat:1997qh,
    author = "Chabanat, E. and Meyer, J. and Bonche, P. and Schaeffer, R. and Haensel, P.",
    title = "{A Skyrme parametrization from subnuclear to neutron star densities}",
    doi = "10.1016/S0375-9474(97)00596-4",
    journal = "Nucl. Phys. A",
    volume = "627",
    pages = "710--746",
    year = "1997"
}

@article{Chabanat:1997un,
    author = "Chabanat, E. and Bonche, P. and Haensel, P. and Meyer, J. and Schaeffer, R.",
    title = "{A Skyrme parametrization from subnuclear to neutron star densities. 2. Nuclei far from stablities}",
    doi = "10.1016/S0375-9474(98)00180-8",
    journal = "Nucl. Phys. A",
    volume = "635",
    pages = "231--256",
    year = "1998",
    note = "[Erratum: Nucl.Phys.A 643, 441--441 (1998)]"
}

@article{Kortelainen:2010hv,
    author = "Kortelainen, M. and Lesinski, T. and More, J. and Nazarewicz, W. and Sarich, J. and Schunck, N. and Stoitsov, M. V. and Wild, S.",
    title = "{Nuclear Energy Density Optimization}",
    eprint = "1005.5145",
    archivePrefix = "arXiv",
    primaryClass = "nucl-th",
    doi = "10.1103/PhysRevC.82.024313",
    journal = "Phys. Rev. C",
    volume = "82",
    pages = "024313",
    year = "2010"
}

@article{Kortelainen:2011ft,
    author = "Kortelainen, M. and McDonnell, J. and Nazarewicz, W. and Reinhard, P. G. and Sarich, J. and Schunck, N. and Stoitsov, M. V. and Wild, S. M.",
    title = "{Nuclear energy density optimization: Large deformations}",
    eprint = "1111.4344",
    archivePrefix = "arXiv",
    primaryClass = "nucl-th",
    doi = "10.1103/PhysRevC.85.024304",
    journal = "Phys. Rev. C",
    volume = "85",
    pages = "024304",
    year = "2012"
}

@article{Kortelainen:2013faa,
    author = "Kortelainen, M. and others",
    title = "{Nuclear energy density optimization: Shell structure}",
    eprint = "1312.1746",
    archivePrefix = "arXiv",
    primaryClass = "nucl-th",
    doi = "10.1103/PhysRevC.89.054314",
    journal = "Phys. Rev. C",
    volume = "89",
    number = "5",
    pages = "054314",
    year = "2014"
}

@article{Qu:2025vib,
    author = "Qu, Xiao-Ying and Chen, Kang-Min and Pan, Cong and Yu, Yang-Yang and Zhang, Kai-Yuan",
    title = "{Benchmarking nuclear energy density functionals with new mass data}",
    eprint = "2505.09914",
    archivePrefix = "arXiv",
    primaryClass = "nucl-th",
    doi = "10.1007/s41365-025-01821-1",
    journal = "Nucl. Sci. Tech.",
    volume = "36",
    number = "12",
    pages = "231",
    year = "2025"
}

@article{Minato:2025ozj,
    author = "Minato, Futoshi and Niu, Yifei and Yoshida, Kenichi",
    title = "{Correlations of Q{\ensuremath{\beta}} values with symmetry energy and effective mass studied within Skyrme energy-density functionals}",
    eprint = "2505.10247",
    archivePrefix = "arXiv",
    primaryClass = "nucl-th",
    doi = "10.1103/c2sz-p3n9",
    journal = "Phys. Rev. C",
    volume = "112",
    number = "4",
    pages = "044314",
    year = "2025"
}

@article{Yoshida:2023zaa,
    author = "Yoshida, Kenichi and Niu, Yifei and Minato, Futoshi",
    title = "{{\ensuremath{\beta}}-decay half-lives as an indicator of shape-phase transition in neutron-rich Zr isotopes with particle-vibration coupling effects}",
    eprint = "2307.00817",
    archivePrefix = "arXiv",
    primaryClass = "nucl-th",
    doi = "10.1103/PhysRevC.108.034305",
    journal = "Phys. Rev. C",
    volume = "108",
    number = "3",
    pages = "034305",
    year = "2023"
}

@article{Kurasawa:2020fli,
    author = "Kurasawa, Haruki and Suda, Toshimi and Suzuki, Toshio",
    title = "{The mean square radius of the neutron distribution and the skin thickness derived from electron scattering}",
    eprint = "2009.00759",
    archivePrefix = "arXiv",
    primaryClass = "nucl-th",
    doi = "10.1093/ptep/ptaa177",
    journal = "PTEP",
    volume = "2021",
    number = "1",
    pages = "013D02",
    year = "2021"
}

@article{Horiuchi:2021dku,
    author = "Horiuchi, Wataru",
    title = "{Single-particle decomposition of nuclear surface diffuseness}",
    eprint = "2110.04982",
    archivePrefix = "arXiv",
    primaryClass = "nucl-th",
    doi = "10.1093/ptep/ptab136",
    journal = "PTEP",
    volume = "2021",
    number = "12",
    pages = "123D01",
    year = "2021"
}

@article{Miyagi:2025lmv,
    author = "Miyagi, Takayuki",
    title = "{Nuclear radii from first principles}",
    doi = "10.3389/fphy.2025.1581854",
    journal = "Front. in Phys.",
    volume = "13",
    pages = "1581854",
    year = "2025"
}

@article{Barrett:2013nh,
    author = "Barrett, Bruce R. and Navratil, Petr and Vary, James P.",
    title = "{Ab initio no core shell model}",
    doi = "10.1016/j.ppnp.2012.10.003",
    journal = "Prog. Part. Nucl. Phys.",
    volume = "69",
    pages = "131--181",
    year = "2013"
}

@article{Launey:2021sua,
    author = "Launey, Kristina D. and Mercenne, Alexis and Dytrych, Tomas",
    title = "{Nuclear Dynamics and Reactions in the Ab Initio Symmetry-Adapted Framework}",
    eprint = "2108.04894",
    archivePrefix = "arXiv",
    primaryClass = "nucl-th",
    doi = "10.1146/annurev-nucl-102419-033316",
    journal = "Ann. Rev. Nucl. Part. Sci.",
    volume = "71",
    pages = "253--277",
    year = "2021"
}

@misc{Sun:2026eep,
    author = "Sun, Xiang-Xiang and Baru, Vadim and Filin, Arseniy A. and Epelbaum, Evgeny and Krebs, Hermann and Mei{\ss}ner, Ulf-G. and Nogga, Andreas",
    title = "{Ab initio charge form factors and radii of light isoscalar nuclei: Role of the two-body charge density}",
    eprint = "2601.09614",
    archivePrefix = "arXiv",
    primaryClass = "nucl-th",
    month = "1",
    year = "2026"
}

@article{Hergert:2015awm,
    author = "Hergert, H. and Bogner, S. K. and Morris, T. D. and Schwenk, A. and Tsukiyama, K.",
    title = "{The In-Medium Similarity Renormalization Group: A Novel Ab Initio Method for Nuclei}",
    eprint = "1512.06956",
    archivePrefix = "arXiv",
    primaryClass = "nucl-th",
    doi = "10.1016/j.physrep.2015.12.007",
    journal = "Phys. Rept.",
    volume = "621",
    pages = "165--222",
    year = "2016"
}

@article{Miyagi:2025rvx,
    author = "Miyagi, T. and Heinz, M. and Schwenk, A.",
    title = "{Ab initio computations of the fourth-order charge density moments of 48Ca and 208Pb}",
    eprint = "2508.10767",
    archivePrefix = "arXiv",
    primaryClass = "nucl-th",
    doi = "10.1016/j.physletb.2025.140032",
    journal = "Phys. Lett. B",
    volume = "872",
    pages = "140032",
    year = "2026"
}

@article{Heinz:2024juw,
    author = "Heinz, M. and Miyagi, T. and Stroberg, S. R. and Tichai, A. and Hebeler, K. and Schwenk, A.",
    title = "{Improved structure of calcium isotopes from ab initio calculations}",
    eprint = "2411.16014",
    archivePrefix = "arXiv",
    primaryClass = "nucl-th",
    doi = "10.1103/PhysRevC.111.034311",
    journal = "Phys. Rev. C",
    volume = "111",
    number = "3",
    pages = "034311",
    year = "2025"
}

@article{Hiyama:2003cu,
    author = "Hiyama, E. and Kino, Y. and Kamimura, M.",
    title = "{Gaussian expansion method for few-body systems}",
    doi = "10.1016/S0146-6410(03)90015-9",
    journal = "Prog. Part. Nucl. Phys.",
    volume = "51",
    pages = "223--307",
    year = "2003"
}

@article{Masui:2014nma,
    author = "Masui, H. and Kat{\={o}}, K. and Michel, N. and P{\l}oszajczak, M.",
    title = "{Precise comparison of the Gaussian expansion method and the Gamow shell model}",
    eprint = "1403.0160",
    archivePrefix = "arXiv",
    primaryClass = "nucl-th",
    doi = "10.1103/PhysRevC.89.044317",
    journal = "Phys. Rev. C",
    volume = "89",
    number = "4",
    pages = "044317",
    year = "2014"
}

@article{Hiyama:2012sma,
    author = "Hiyama, Emiko",
    title = "{Gaussian expansion method for few-body systems and its applications to atomic and nuclear physics}",
    doi = "10.1093/ptep/pts015",
    journal = "PTEP",
    volume = "2012",
    pages = "01A204",
    year = "2012"
}

@article{Burrows:2017wqn,
    author = "Burrows, M. and Elster, Ch. and Popa, G. and Launey, K. D. and Nogga, A. and Maris, P.",
    title = "{Ab initio Translationally Invariant Nonlocal One-body Densities from No-core Shell-model Theory}",
    eprint = "1711.07080",
    archivePrefix = "arXiv",
    primaryClass = "nucl-th",
    doi = "10.1103/PhysRevC.97.024325",
    journal = "Phys. Rev. C",
    volume = "97",
    number = "2",
    pages = "024325",
    year = "2018"
}

@article{Burrows:2018ggt,
    author = "Burrows, M. and Elster, Ch. and Weppner, S. P. and Launey, K. D. and Maris, P. and Nogga, A. and Popa, G.",
    title = "{Ab initio folding potentials for nucleon-nucleus scattering based on no-core shell-model one-body densities}",
    eprint = "1810.06442",
    archivePrefix = "arXiv",
    primaryClass = "nucl-th",
    doi = "10.1103/PhysRevC.99.044603",
    journal = "Phys. Rev. C",
    volume = "99",
    number = "4",
    pages = "044603",
    year = "2019"
}

@article{Foy:2025yot,
    author = "Foy, J. and Elster, Ch. and Maris, P. and Weppner, S. P. and Bogner, S. K.",
    title = "{Separable character of ab initio no-core shell model one-body densities}",
    eprint = "2508.09890",
    archivePrefix = "arXiv",
    primaryClass = "nucl-th",
    doi = "10.1103/59tv-8c7d",
    journal = "Phys. Rev. C",
    volume = "113",
    number = "4",
    pages = "044323",
    year = "2026"
}

@article{Sun:2025yfo,
    author = "Sun, Xiang-Xiang and Le, Hoai and Mei{\ss}ner, Ulf-G. and Nogga, Andreas",
    title = "{Radii of light nuclei from the Jacobi no-core shell model}",
    eprint = "2502.03989",
    archivePrefix = "arXiv",
    primaryClass = "nucl-th",
    doi = "10.1103/j4ky-tn1j",
    journal = "Phys. Rev. C",
    volume = "112",
    number = "2",
    pages = "024317",
    year = "2025"
}

@article{Hascoet:2025zfm,
    author = {Hasco{\"e}t, Laurent and Menickelly, Matt and Narayanan, Sri Hari Krishna and O'Neal, Jared and Schunck, Nicolas and Wild, Stefan M.},
    title = "{HFBTHO-AD: Differentiation of a nuclear energy density functional code}",
    eprint = "2508.11910",
    archivePrefix = "arXiv",
    primaryClass = "nucl-th",
    doi = "10.1016/j.cpc.2025.109955",
    journal = "Comput. Phys. Commun.",
    volume = "320",
    pages = "109955",
    year = "2026"
}

@misc{Yoshimura:2026cdk,
    author = "Yoshimura, Kenta",
    title = "{Neural-Network-Based Variational Method in Nuclear Density Functional Theory: Application to the Extended Thomas-Fermi Model}",
    eprint = "2604.25759",
    archivePrefix = "arXiv",
    primaryClass = "nucl-th",
    month = "4",
    year = "2026"
}

@article{Scamps:2020fyu,
    author = "Scamps, Guillaume and Goriely, Stephane and Olsen, Erik and Bender, Michael and Ryssens, Wouter",
    title = "{Skyrme-Hartree-Fock-Bogoliubov mass models on a 3D mesh: effect of triaxial shape}",
    eprint = "2011.07904",
    archivePrefix = "arXiv",
    primaryClass = "nucl-th",
    doi = "10.1140/epja/s10050-021-00642-1",
    journal = "Eur. Phys. J. A",
    volume = "57",
    number = "12",
    pages = "333",
    year = "2021"
}

@article{Agrawal:2005ix,
    author = "Agrawal, B. K. and Shlomo, S. and Au, V. Kim",
    title = "{Determination of the parameters of a Skyrme type effective interaction using the simulated annealing approach}",
    eprint = "nucl-th/0505071",
    archivePrefix = "arXiv",
    doi = "10.1103/PhysRevC.72.014310",
    journal = "Phys. Rev. C",
    volume = "72",
    pages = "014310",
    year = "2005"
}

@article{Zhang:2015vaa,
    author = "Zhang, Zhen and Chen, Lie-Wen",
    title = "{Extended Skyrme interactions for nuclear matter, finite nuclei and neutron stars}",
    eprint = "1510.06459",
    archivePrefix = "arXiv",
    primaryClass = "nucl-th",
    doi = "10.1103/PhysRevC.94.064326",
    journal = "Phys. Rev. C",
    volume = "94",
    number = "6",
    pages = "064326",
    year = "2016"
}

@article{Amiri:2026zbz,
    author = "Amiri, M. M. and Ghodsi, O. N.",
    title = "{Toward a dedicated Skyrme interaction for {\ensuremath{\alpha}} decay of superheavy nuclei}",
    doi = "10.1103/353g-7f4s",
    journal = "Phys. Rev. C",
    volume = "114",
    number = "1",
    pages = "014601",
    year = "2026"
}

@article{Phillips:2020dmw,
    author = "Phillips, D. R. and others",
    title = "{Get on the BAND Wagon: A Bayesian Framework for Quantifying Model Uncertainties in Nuclear Dynamics}",
    eprint = "2012.07704",
    archivePrefix = "arXiv",
    primaryClass = "nucl-th",
    doi = "10.1088/1361-6471/abf1df",
    journal = "J. Phys. G",
    volume = "48",
    number = "7",
    pages = "072001",
    year = "2021"
}

@article{Burgio:2021vgk,
    author = "Burgio, G. F. and Schulze, H. -J. and Vidana, I. and Wei, J. -B.",
    title = "{Neutron stars and the nuclear equation of state}",
    eprint = "2105.03747",
    archivePrefix = "arXiv",
    primaryClass = "nucl-th",
    doi = "10.1016/j.ppnp.2021.103879",
    journal = "Prog. Part. Nucl. Phys.",
    volume = "120",
    pages = "103879",
    year = "2021"
}

@article{Semposki:2025etb,
    author = "Semposki, A. C. and Drischler, C. and Furnstahl, R. J. and Phillips, D. R.",
    title = "{Microscopic constraints for the equation~of state and structure of neutron stars: A Bayesian model mixing framework}",
    eprint = "2505.18921",
    archivePrefix = "arXiv",
    primaryClass = "nucl-th",
    doi = "10.1103/fxv6-gdnw",
    journal = "Phys. Rev. C",
    volume = "113",
    number = "1",
    pages = "015808",
    year = "2026"
}

@article{Miller:2019cac,
    author = "Miller, M. C. and others",
    title = "{PSR J0030+0451 Mass and Radius from $NICER$ Data and Implications for the Properties of Neutron Star Matter}",
    eprint = "1912.05705",
    archivePrefix = "arXiv",
    primaryClass = "astro-ph.HE",
    doi = "10.3847/2041-8213/ab50c5",
    journal = "Astrophys. J. Lett.",
    volume = "887",
    number = "1",
    pages = "L24",
    year = "2019"
}

@article{Pal:2026cji,
    author = "Pal, Suman and Chaudhuri, Gargi",
    title = "{Characterizing the quark-hadron mixed phase in compact star cores: Sensitivity to nuclear saturation and quark-model parameters at finite temperature}",
    eprint = "2605.05005",
    archivePrefix = "arXiv",
    primaryClass = "nucl-th",
    doi = "10.1103/8gqy-7ym2",
    journal = "Phys. Rev. D",
    volume = "113",
    number = "10",
    pages = "103006",
    year = "2026"
}

@article{Chamel:2025fhb,
    author = "Chamel, Nicolas",
    title = "{Superfluid fraction in the crystalline crust of a neutron star: Role of quantum zero-point motion of ions}",
    eprint = "2605.20934",
    archivePrefix = "arXiv",
    primaryClass = "astro-ph.HE",
    doi = "10.1103/PhysRevC.111.055803",
    journal = "Phys. Rev. C",
    volume = "111",
    number = "5",
    pages = "055803",
    year = "2025"
}

@article{Nakatsukasa2026,
author = {Nakatsukasa, Takashi and Hinohara, Nobuo},
year = {2026},
month = {07},
pages = {},
title = {The finite amplitude method for nuclear collective motion},
volume = {62},
journal = {The European Physical Journal A},
doi = {10.1140/epja/s10050-026-01862-z}
}

@article{Klausner:2026foh,
    author = "Klausner, Pietro and Antonelli, Marco and Col{\`o}, Gianluca and Gulminelli, Francesca and Roca-Maza, Xavier and Vigezzi, Enrico",
    title = "{Emulator-assisted nuclear density-functional-theory inference and its consequences for the structure of neutron stars}",
    eprint = "2604.11358",
    archivePrefix = "arXiv",
    primaryClass = "nucl-th",
    doi = "10.1103/p6xt-x2zp",
    journal = "Phys. Rev. C",
    volume = "114",
    number = "3",
    pages = "034320",
    year = "2026"
}

@article{Klausner:2025ucq,
    author = "Klausner, Pietro and Antonelli, Marco and Gulminelli, Francesca",
    title = "{Properties of the neutron star crust informed by nuclear structure data}",
    eprint = "2505.16929",
    archivePrefix = "arXiv",
    primaryClass = "nucl-th",
    doi = "10.1103/mm6h-3jqs",
    journal = "Phys. Rev. C",
    volume = "113",
    number = "2",
    pages = "025808",
    year = "2026"
}

@article{Klausner:2024jgu,
    author = "Klausner, Pietro and Col{\`o}, Gianluca and Roca-Maza, Xavier and Vigezzi, Enrico",
    title = "{Impact of ground-state properties and collective excitations on the Skyrme ansatz: A Bayesian study}",
    eprint = "2410.18598",
    archivePrefix = "arXiv",
    primaryClass = "nucl-th",
    doi = "10.1103/PhysRevC.111.014311",
    journal = "Phys. Rev. C",
    volume = "111",
    number = "1",
    pages = "014311",
    year = "2025"
}

@article{Sun:2023xkg,
    author = "Sun, Boyang and Bhattiprolu, Saketh and Lattimer, James M.",
    title = "{Compiled properties of nucleonic matter and nuclear and neutron star models from nonrelativistic and relativistic interactions}",
    eprint = "2311.00843",
    archivePrefix = "arXiv",
    primaryClass = "nucl-th",
    doi = "10.1103/PhysRevC.109.055801",
    journal = "Phys. Rev. C",
    volume = "109",
    number = "5",
    pages = "055801",
    year = "2024"
}

@article{Neufcourt:2018syo,
    author = "Neufcourt, L{\'e}o and Cao, Yuchen and Nazarewicz, Witold and Viens, Frederi",
    title = "{Bayesian approach to model-based extrapolation of nuclear observables}",
    eprint = "1806.00552",
    archivePrefix = "arXiv",
    primaryClass = "nucl-th",
    doi = "10.1103/PhysRevC.98.034318",
    journal = "Phys. Rev. C",
    volume = "98",
    number = "3",
    pages = "034318",
    year = "2018"
}

@article{Xie:2026rep,
    author = "Xie, Wen-Jie and Xia, Cheng-Jun",
    title = "{Impact of radius-measurement uncertainty on RMF parameters and neutron-star matter properties}",
    doi = "10.1007/s41365-026-01988-1",
    journal = "Nucl. Sci. Tech.",
    volume = "37",
    number = "9",
    pages = "161",
    year = "2026"
}

@article{Li:2025oxi,
    author = "Li, Jia-Jie and Tian, Yu and Sedrakian, Armen",
    title = "{Bayesian inferences on covariant density functionals from multimessenger astrophysical data: Nucleonic models}",
    eprint = "2502.20000",
    archivePrefix = "arXiv",
    primaryClass = "nucl-th",
    doi = "10.1103/PhysRevC.111.055804",
    journal = "Phys. Rev. C",
    volume = "111",
    number = "5",
    pages = "055804",
    year = "2025"
}

@article{Wei:2025aku,
    author = "Wei, Guo-Jun and Li, Jia-Jie and Sedrakian, Armen and Wang, Yong-Jia and Li, Qing-Feng and Liu, Fu-Hu",
    title = "{Bayesian inferences on covariant density functionals from multimessenger astrophysical data: Influences of parametrizations of density-dependent couplings}",
    eprint = "2512.01503",
    archivePrefix = "arXiv",
    primaryClass = "astro-ph.HE",
    doi = "10.1103/n336-1sqq",
    journal = "Phys. Rev. C",
    volume = "113",
    number = "5",
    pages = "055805",
    year = "2026"
}

@misc{Aoyama:2026fbd,
    author = "Aoyama, Shigeyoshi",
    title = "{Bayesian Variational Method for Precision Few-Body Calculations}",
    eprint = "2607.25265",
    archivePrefix = "arXiv",
    primaryClass = "nucl-th",
    month = "7",
    year = "2026"
}

@article{Choudhury:2024xbk,
    author = "Choudhury, Devarshi and others",
    title = "{A NICER View of the Nearest and Brightest Millisecond Pulsar: PSR J0437{\textendash}4715}",
    eprint = "2407.06789",
    archivePrefix = "arXiv",
    primaryClass = "astro-ph.HE",
    doi = "10.3847/2041-8213/ad5a6f",
    journal = "Astrophys. J. Lett.",
    volume = "971",
    number = "1",
    pages = "L20",
    year = "2024"
}

@article{Ozel:2016oaf,
    author = {{\"O}zel, Feryal and Freire, Paulo},
    title = "{Masses, Radii, and the Equation of State of Neutron Stars}",
    eprint = "1603.02698",
    archivePrefix = "arXiv",
    primaryClass = "astro-ph.HE",
    doi = "10.1146/annurev-astro-081915-023322",
    journal = "Ann. Rev. Astron. Astrophys.",
    volume = "54",
    pages = "401--440",
    year = "2016"
}

@article{Fracasso:2007fi,
    author = "Fracasso, Sara and Colo, Gianluca",
    title = "{Spin-isospin nuclear response using the existing microscopic Skyrme functionals}",
    eprint = "0704.2892",
    archivePrefix = "arXiv",
    primaryClass = "nucl-th",
    doi = "10.1103/PhysRevC.76.044307",
    journal = "Phys. Rev. C",
    volume = "76",
    pages = "044307",
    year = "2007"
}

@article{Bennaceur:2005mx,
    author = "Bennaceur, K. and Dobaczewski, J.",
    title = "{Coordinate-space solution of the Skyrme-Hartree-Fock-Bogolyubov equations within spherical symmetry. The Program HFBRAD (v1.00)}",
    eprint = "nucl-th/0501002",
    archivePrefix = "arXiv",
    doi = "10.1016/j.cpc.2005.02.002",
    journal = "Comput. Phys. Commun.",
    volume = "168",
    pages = "96--122",
    year = "2005"
}

@article{Chen:2009wv,
    author = "Chen, Lie-Wen and Cai, Bao-Jun and Ko, Che Ming and Li, Bao-An and Shen, Chun and Xu, Jun",
    title = "{High-order effects on the incompressibility of isospin asymmetric nuclear matter}",
    eprint = "0905.4323",
    archivePrefix = "arXiv",
    primaryClass = "nucl-th",
    doi = "10.1103/PhysRevC.80.014322",
    journal = "Phys. Rev. C",
    volume = "80",
    pages = "014322",
    year = "2009"
}

@article{Chen:2010qx,
    author = "Chen, Lie-Wen and Ko, Che Ming and Li, Bao-An and Xu, Jun",
    title = "{Density slope of the nuclear symmetry energy from the neutron skin thickness of heavy nuclei}",
    eprint = "1004.4672",
    archivePrefix = "arXiv",
    primaryClass = "nucl-th",
    doi = "10.1103/PhysRevC.82.024321",
    journal = "Phys. Rev. C",
    volume = "82",
    pages = "024321",
    year = "2010"
}

@article{Zhao:2022xhq,
    author = "Zhao, Qiang and Ren, Zhengxue and Zhao, Pengwei and Meng, Jie",
    title = "{Covariant density functional theory with localized exchange terms}",
    eprint = "2207.01764",
    archivePrefix = "arXiv",
    primaryClass = "nucl-th",
    doi = "10.1103/PhysRevC.106.034315",
    journal = "Phys. Rev. C",
    volume = "106",
    number = "3",
    pages = "034315",
    year = "2022"
}

@article{Roth:2007sv,
    author = "Roth, R. and Navratil, P.",
    title = "{Ab initio study of Ca-40 with an importance truncated no-core shell model}",
    eprint = "0705.4069",
    archivePrefix = "arXiv",
    primaryClass = "nucl-th",
    reportNumber = "UCRL-JRNL-231224",
    doi = "10.1103/PhysRevLett.99.092501",
    journal = "Phys. Rev. Lett.",
    volume = "99",
    pages = "092501",
    year = "2007"
}

@article{corner,
  doi = {10.21105/joss.00024},
  url = {https://doi.org/10.21105/joss.00024},
  year  = {2016},
  month = {jun},
  publisher = {The Open Journal},
  volume = {1},
  number = {2},
  pages = {24},
  author = {Daniel Foreman-Mackey},
  title = {corner.py: Scatterplot matrices in Python},
  journal = {The Journal of Open Source Software}
}

@article{Gil:2020wqs,
    author = "Gil, Hana and Kim, Young-Min and Papakonstantinou, Panagiota and Hyun, Chang Ho",
    title = "{Constraining the density dependence of the symmetry energy with nuclear data and astronomical observations in the Korea-IBS-Daegu-SKKU framework}",
    eprint = "2010.13354",
    archivePrefix = "arXiv",
    primaryClass = "nucl-th",
    doi = "10.1103/PhysRevC.103.034330",
    journal = "Phys. Rev. C",
    volume = "103",
    number = "3",
    pages = "034330",
    year = "2021"
}
%%%%%%%%%%%%%%%%%%%%%%%%%%%%%%%%%%%%%%%%%%%%%%%%%

\end{document}